\documentclass[STIX1COL]{WileyNJD-v2}

\articletype{Article Type}%

\received{26 April 2016}
\revised{6 June 2016}
\accepted{6 June 2016}

\usepackage{graphicx}          %
\usepackage{bm}
\usepackage{graphicx}
\usepackage{hyperref}
\usepackage{dsfont}
\usepackage{enumitem} 
\usepackage{arydshln} 

\usepackage{irt-common}
\usepackage{irt-shortcuts}
\usepackage{irt-tikz}
\usetikzlibrary{patterns}
\usepackage{nicematrix}

\newcommand{\Ts}{\ensuremath{\T_{\mathrm{s}}}}
\newcommand{\To}{\ensuremath{\T_{\mathrm{o}}}}
\newcommand{\Ti}{\ensuremath{\T_{\mathrm{i}}}}
\newcommand{\Td}{\ensuremath{\T_{\mathrm{d}}}}
\newcommand{\Aa}{\ensuremath{\A_{\mathrm{a}}}}
\newcommand{\Ab}{\ensuremath{\A_{\mathrm{b}}}}
\newcommand{\Ac}{\ensuremath{\A_{\mathrm{c}}}}
\newcommand{\Ad}{\ensuremath{\A_{\mathrm{d}}}}
\newcommand{\Adbar}{\ensuremath{\bar\A_{\mathrm{d}}}}
\newcommand{\Adbarcheck}{\ensuremath{\check{\bar{\A}}_{\mathrm{d}}}}
																	
\newcommand{\Cb}{\ensuremath{\C_{\mathrm{b}}}}
\newcommand{\Cd}{\ensuremath{\C_{\mathrm{d}}}}
\newcommand{\Bc}{\ensuremath{\B_{\mathrm{c}}}}
\newcommand{\Bd}{\ensuremath{\B_{\mathrm{d}}}}
\newcommand{\Hdd}{\ensuremath{\H_{\mathrm{dd}}}}
\newcommand{\Hddbar}{\ensuremath{\bar\H_{\mathrm{dd}}}}
\newcommand{\Eca}{\ensuremath{\F_{\mathrm{ca}}}}
\newcommand{\Hcb}{\ensuremath{\H_{\mathrm{cb}}}}
\newcommand{\Hcd}{\ensuremath{\H_{\mathrm{cd}}}}
\newcommand{\Lb}{\ensuremath{\L_{\mathrm{b}}}}
\newcommand{\Fdc}{\ensuremath{\F_{\mathrm{dc}}}}
\newcommand{\Fda}{\ensuremath{\F_{\mathrm{da}}}}
\newcommand{\Fdb}{\ensuremath{\F_{\mathrm{db}}}}
\newcommand{\Fdd}{\ensuremath{\F_{\mathrm{dd}}}}
\newcommand{\Fddbar}{\ensuremath{\bar\F_{\mathrm{dd}}}}
\newcommand{\Hab}{\ensuremath{\H_{\mathrm{ab}}}}
\newcommand{\Had}{\ensuremath{\H_{\mathrm{ad}}}}
\newcommand{\Hbd}{\ensuremath{\H_{\mathrm{bd}}}}
\newcommand{\Hbb}{\ensuremath{\H_{\mathrm{bb}}}}
\newcommand{\Boa}{\ensuremath{\B_{0\mathrm{a}}}}
\newcommand{\Bob}{\ensuremath{\B_{0\mathrm{b}}}}
\newcommand{\Boc}{\ensuremath{\B_{0\mathrm{c}}}}
\newcommand{\Bod}{\ensuremath{\B_{0\mathrm{d}}}}

\newcommand{\Coa}{\ensuremath{\C_{0\mathrm{a}}}}
\newcommand{\Cob}{\ensuremath{\C_{0\mathrm{b}}}}
\newcommand{\Coc}{\ensuremath{\C_{0\mathrm{c}}}}
\newcommand{\Cod}{\ensuremath{\C_{0\mathrm{d}}}}

\newcommand{\ud}{\ensuremath{\u_{\mathrm{d}}}}
\newcommand{\uc}{\ensuremath{\u_{\mathrm{c}}}}
\newcommand{\yd}{\ensuremath{\y_{\mathrm{d}}}}
\newcommand{\yb}{\ensuremath{\y_{\mathrm{b}}}}
\newcommand{\lb}{\ensuremath{\l_{\mathrm{b}}}}
\newcommand{\ld}{\ensuremath{\l_{\mathrm{d}}}}
\newcommand{\xa}{\ensuremath{\x_{\mathrm{a}}}}
\newcommand{\na}{\ensuremath{n_\mathrm{a}}}
\newcommand{\nb}{\ensuremath{n_\mathrm{b}}}
\newcommand{\nc}{\ensuremath{n_\mathrm{c}}}
\newcommand{\nd}{\ensuremath{n_\mathrm{d}}}
\newcommand{\md}{\ensuremath{m_{\mathrm{d}}}}
\newcommand{\mc}{\ensuremath{m_{\mathrm{c}}}}
\newcommand{\mnull}{\ensuremath{m_{{0}}}}
\newcommand{\pb}{\ensuremath{p_{\mathrm{b}}}}

\theoremstyle{definition}

\begin{document}

\title{Unknown Input Observer Design for Linear Time-Invariant Systems -- A Unifying Framework\protect\thanks{
This work was supported by the Graz University of Technology LEAD project ``Dependable Internet of Things in Adverse Environments'' and by the European Union’s Horizon 2020 research and innovation programme under the Marie Sklodowska-Curie Grant agreement 734832. The financial support by the Christian Doppler Research Association, Austria, the Austrian Federal Ministry for Digital and Economic Affairs and the National Foundation for Research, Technology and Development, Austria is gratefully acknowledged.
}}

\author[1]{Markus Tranninger*}
\author[1,2]{Helmut Niederwieser}
\author[1,3]{Richard Seeber}
\author[1,3]{Martin Horn}

\authormark{M. Tranninger, H. Niederwieser and R. Seeber}

\address[1]{\orgdiv{Institute of Automation and Control}, \orgname{Graz University of Technology}, \orgaddress{
		\state{Graz},
		 \country{Austria}}}

\address[2]{\orgname{BEST -- Bioenergy and Sustainable Technologies GmbH}, \orgaddress{\state{Graz}, \country{Austria}}}

\address[3]{\orgdiv{Christian Doppler Laboratory for Model-Based Control of Complex Test Bed Systems, Institute of Automation and Control}, \orgname{Graz University of Technology}, \orgaddress{\state{Graz}, \country{Austria}}}

\corres{*Markus Tranninger, \email{markus.tranninger@tugraz.at}}

\abstract[Summary]{
	This paper presents a new observer design approach for linear time invariant multivariable systems subject to unknown inputs.
	The design is based on a transformation to the so-called special coordinate basis.
	This form reveals important system properties like invertability or the finite and infinite zero structure.
	Depending on the system's strong observability properties, the special coordinate basis allows for a straightforward unknown input observer design utilizing linear or nonlinear observers design techniques.
	The chosen observer design technique does not only depend on the system properties, but also on the desired convergence behavior of the observer.
	Hence, the proposed design procedure can be seen as a unifying framework for unknown input observer design.
}


\maketitle

\section{Introduction}\label{sec:introduction}

State estimation in the presence of unknown inputs has a long history in research~\cite{bhattacharyya1978observer,molinari1976astrong,hautus1983strong,hou1994fault,kratz1995characterization,valcher1999state} and applications~\cite{chen1999robust,alwi2008fault}.
So-called unknown input observers can provide state estimates even if not all system inputs are known.
During the last decades, such observers proved useful in a large variety of applications ranging from robust control and uncertainty compensation~\cite{ferreira2011robust} over robust residual generation in fault detection problems~\cite{chen1999robust} to networked and decentralized control scenarios~\cite{saif1992decentralized,taha2015unknown}.

The majority of the available works deals with the design of linear unknown input observers for linear time-invariant systems, see, e.g.~\cite{hautus1983strong,valcher1999state,hou1994fault,saif1992decentralized,taha2015unknown}.
The necessary and sufficient existence condition for such an observer is the system's strong$^*$ detectability as introduced by Hautus~\cite{hautus1983strong}.
Such systems do not possess unstable invariant zeros and have to fulfill the so-called rank condition. 
The latter condition, also known as observer matching condition, allows to express the unknown input via the first derivative of the output signal~\cite{hou1994fault}.

Strong$^*$ detectability is also necessary and sufficient for the existence of first order sliding mode observers, see~\cite[Chapter 6]{edwards1998sliding}.
The advantage of such observers compared to linear ones is that they can also provide estimates of the unknown input via an equivalent control approach.
A comparison between linear and sliding mode observers for fault reconstruction is presented in~\cite{edwards2006acomparison}.
Such first order sliding mode techniques were also successfully applied to, e.g., fault detection and fault tolerant control problems in aviation~\cite{alwi2008fault,alwi2011fault}.

If the system is strongly detectable~\cite{hautus1983strong} but the observer matching condition is not fulfilled, it is still possible to estimate the states by taking higher order derivatives of the output signal into account~\cite{tranninger2019exact}.
Popular techniques for obtaining such derivatives are based on higher order sliding mode techniques and in particular on Levant's arbitrary order robust exact differentiator~\cite{levant1998robust}.
This differentiator was recently generalized by J. Moreno to a family of finite- or fixed-time bi-homogeneous differentiators~\cite{moreno2020arbitrary}.

Concerning sliding mode based unknown input observers, the early works proposed a hierarchical super-twisting algorithm~\cite{bejarano2007exact}. 
The underlying idea is Molinari's algorithm~\cite{molinari1976astrong} for determining the weakly unobservable subspace. It is based on successive differentiation of the output vector and the decoupling from the influence of the unknown inputs.
The approach was extended in order to employ higher order sliding mode differentiators~\cite{tranninger2019exact,bejarano2009unknown,fridman2011high-order,bejarano2010high-order}.
This improves the accuracy of the obtained estimates w.r.t. discretization and bounded measurement noise~\cite{ferreira2011robust}. 
All mentioned higher order sliding mode approaches require a Luenberger observer cascaded with the sliding mode reconstruction scheme in order to fulfill the requirements for the higher order sliding mode differentiators~\cite{tranninger2019exact}. 
Hence, such estimation schemes require at least twice the number of states of the considered system.
For strongly observable single-input single-output (SISO) systems, a direct higher order sliding mode observer design which avoids this disadvantage and requires less tuning parameters is presented in~\cite{niederwieser2019ageneralization}. 
The design is based on a generalization of Ackermann's formula. 
Recently, a generalization to strongly observable multi-input multi-output (MIMO) systems was proposed in~\cite{niederwieser2021high}. 
The design is based on a new observer normal form for multivariable systems.
This normal form allows a direct application of the robust exact differentiator (RED) without a cascaded Luenberger observer and hence simplifies the design procedure and the tuning compared to previous approaches.

A generalization of the above normal form is the so-called special coordinate basis (SCB), which explicitly reveals the finite and infinite zero structure of the considered system and was introduced by Sannuti and Saberi in~\cite{sannuti1987special}.
Since then, the SCB is utilized to solve many analysis and design problems for multivariable linear time-invariant systems, like, e.g., determining invariant subspaces from the geometric control perspective~\cite{chen2004linear}, squaring down~\cite{saberi1990squaring}, model order reduction~\cite{ozcetin1990special}, loop transfer recovery~\cite{chen2011loop}, $H_2$ and $H_\infty$ optimal control~\cite{chen1993construction,chen2000robust} and many more.
A numerically reliable algorithm for obtaining the SCB transformations is proposed in~\cite{chu2002onthenumerical}.
Xiong and Saif were among the first to utilize the special coordinate basis for unknown input observer design ~\cite{xiong1999functional,xiong1999robust,saif2003sliding}. 
In~\cite{xiong1999functional}, they propose a linear functional unknown input observer. 
A robust linear fault isolation observer for strong$^*$ detectable systems is presented in~\cite{xiong1999robust}.
In~\cite{saif2003sliding}, a first order sliding mode observer design technique for fault diagnosis based on the equivalent control principle is proposed for strongly detectable systems. 
Due to the relation of the SCB with the system's zero structure, this form is particularly suitable for the unknown input observer design.

This paper presents a novel observer design approach for linear time invariant MIMO systems subject to unknown inputs.
The design is based on a transformation to the special coordinate basis~\cite{sannuti1987special,chen2004linear}.
In this form, and depending on the system's observability properties, the design of a linear unknown input observer and the design of first order or higher order sliding mode observers can be performed in a similar fashion. 

For linear unknown input observers it is shown that a full order observer has no benefit over a specific reduced order observer.
Moreover, it turns out that the observer rank condition is a condition on the system's infinite zero structure that only allows infinite zeros of degree one.

If the system is merely strongly detectable but not strongly observable, it is still possible to asymptotically estimate the system states by utilizing higher order sliding mode techniques.
It follows directly from the construction of the proposed observer that the number of differentiation operations is minimal.
The design is straightforward and does not require a ``stabilizing'' Luenberger observer as in previous works~\cite{tranninger2019exact,bejarano2009unknown,fridman2011high-order}.
Hence, its observer order corresponds to the system order.

For strongly observable systems, it is well known that it is possible to reconstruct the states in finite time~\cite{bejarano2007exact}.
The present work proposes a finite- or fixed-time unknown input observer design for strongly observable systems.
To that end, depending on whether the observer matching condition is fulfilled or not, a continuous or discontinuous (bi)-homogenous observer based on Moreno's differentiator is proposed, respectively.
To sum up, the proposed design procedure can be regarded as a unifying unknown input observer design framework for linear time invariant systems.
It incorporates the design of linear unknown input observers, sliding mode observers and bi-homogeneous observers allowing for asymptotic, finite- or fixed-time convergence of the estimation error.

The paper is structured as follows: Section~\ref{sec:problemstatement} introduces the problem statement and points out the underlying assumptions.
Section~\ref{sec:preliminaries} recalls preliminaries such as important properties of linear time invariant systems.
Moreover, it summarizes Moreno's arbitrary order fixed-time differentiator and discusses Levant's robust exact differentiator as a special case.
The transformation to the special coordinate basis is introduced in Section~\ref{sec:SCB}.
In order to utilize the SCB for observer design, a specific block triangular form of the subsystem related to the infinite zero structure is proposed.
The existence of such a transformation is guaranteed by Theorem~\ref{THM:TRANSF_SUBSYS_D}.
The main contribution, i.e., the unifying observer design framework, is presented in Section~\ref{sec:observerdesign}.
There, the general structure of the observers is discussed and the particular designs are presented in detail depending on the system's observability properties.
Section~\ref{sec:example} exemplarily shows a sliding mode observer design for a specific strongly detectable system. 
The numerical results underline the applicability and the appealing simplicity of the proposed design procedure.
The conclusion together with possible future research directions is given in Section~\ref{sec:discussion}.

\paragraph{Notation:}
Matrices are printed in boldface capital letters, whereas column vectors are boldface lower case letters. 
The elements of a matrix $\M$ are denoted by $m_{i,j}$.
The matrix $\I_n$ is the $n\times n$ identity matrix and $\J_n$ denotes a Jordan block of dimension $n\times n$ according to
\begin{equation}
	\J_n = \begin{bmatrix}
		\bf 0 & \I_{n-1} \\
		 0 & \bf 0^\transp
	\end{bmatrix}.
\end{equation}
Moreover, $\M= \diag{\M_1,\ldots,\M_j}$ denotes a (block) diagonal matrix with entries $\M_1,\ldots,\M_j$.
In dynamical systems, differentiation of a vector $\x$ with respect to time $t$ is expressed as $\dot \x$.
Time dependency of state (usually $\x$), input (usually $\u$) and output (usually $\y$) is omitted.
The signed power is represented by $\lfloor x \rceil^a=\sign{x}|x|^{a}$ and $\lfloor x \rceil^0 = \sign{x}$. 
For differential equations with a discontinuous right hand side, the solutions are understood in the sense of Filippov~\cite{filippov1988differential}.

\section{Problem Statement}\label{sec:problemstatement}

This paper considers the linear time invariant MIMO system $\Sigma$ denoted by the quadruple $(\A,\B,\C,\D)$ and given by
\begin{subequations}\label{eq:sys:orig}
	\begin{align}
		\dot \x &= \A\x + \B\u,\quad\x(0) = \x_0, \\
		\y &= \C\x + \D \u
	\end{align}
\end{subequations}
with the state $\x\in \mathds R^n$, the unknown input $\u\in\mathds R^m$ and the output $\y \in \mathds R^p$.
Without loss of generality, it is assumed that $\rank\D =m_0$ and that the matrices $\left[\B^\transp \;\D^\transp\right]$ and $\left[\C\;\D\right]$ have full rank.
Moreover, only unknown inputs are considered for simplicity. 
This is no restriction, because known inputs can always be easily integrated in any estimation scheme~\cite{bejarano2009unknown}.

The unknown input $\u = \left[u_1\; u_2\; \cdots \; u_m \right]^\transp$ is assumed to be bounded component-wise according to
\begin{equation}\label{eq:uibound}
u_i \in [u_{i,\min},\; u_{i,\max}],\; i=1,\ldots,m.	
\end{equation}

The concept of an unknown input observer is introduced in
\begin{definition}
An unknown input observer is a (dynamical) system providing an estimate $\hat\x$ for the system state $\x$ without knowledge of the input $\u$. 
Moreover
\begin{enumerate}[label={\roman*)}]
	\item It is an \emph{asymptotic unknown input observer} if the estimation error $\e = \x-\hat \x$ vanishes asymptotically for any initial condition, i.e., $\lim_{t\rightarrow \infty} \e(t) = \bm 0$.
	\item It is a \emph{finite-time unknown input observer}, if $\e(t) = \bm 0$ for all $t\geq T_f(\e_0)$ with some finite time $T_f>0$ depending on the initial error $\e_0$.
	\item It is a \emph{fixed-time unknown input observer} if $T_f$ is independent of $\e_0$.  
\end{enumerate}
\end{definition}

The goal is to derive a generic design procedure which, depending on the system properties, allows to design asymptotic, finite-time or fixed-time unknown input observers whose order is at most the order of the system.

\section{Preliminaries}\label{sec:preliminaries}

In the following, important concepts required for the unknown input observer design are briefly recalled.
Most of the results presented in Sections~\ref{sec:zeros} and~\ref{sec:strongdetectability} can be found in classical textbooks on multivariable control systems~\cite{chen2004linear,trentelman2012control,skogestad2005multivariable}.
Section~\ref{sec:morenodiff} recalls a family of differentiators recently proposed in~\cite{moreno2020arbitrary}.

\subsection{Zeros}\label{sec:zeros}

Conditions for strong detectability and observability are often stated in terms of invariant zeros of system $\Sigma$.
These zeros are characterized by the so-called Rosenbrock matrix, see, e.g.,~\cite[Chapter 7]{trentelman2012control}. 
\begin{definition}[invariant zeros]\label{def:invariantzero}
	The invariant zeros of $\Sigma$ are the values $\lambda\in\mathds C$ such that the Rosenbrock matrix
	\begin{equation}
		\P(s) = \begin{bmatrix} s\I_n-\A & -\B \\ 
			\C & \phantom{-}\D \end{bmatrix}
	\end{equation}
	exhibits a rank loss, i.e.,
	\begin{equation}
		\rank \P(\lambda) < \normrank \P,
	\end{equation}
where the normal rank of $\P(s)$ is defined as
\begin{equation}
	\normrank \P= \max\{\rank\P(s)\;| \;s\in \mathds C\}.
\end{equation}
\end{definition}
It holds, that $\rank \P(s) = n+\normrank{\G(s)}$, with the transfer matrix
\begin{equation}
	\G(s) = \C(s\I_n -\A)^{-1}\B + \D,
\end{equation}
see~\cite[Lemma 8.9]{trentelman2012control}.
Note that this definition does not reveal the multiplicity or order of the zeros. 
In~\cite[Def. 3.6.3]{chen2004linear}, the orders are introduced via the Kronecker canonical form of $\P(s)$. 
There, also the definition of the infinite zero structure is stated. 
This definition is omitted here and the infinite zero structure will be introduced with the aid of the special coordinate basis in Section~\ref{sec:SCB}.
If all invariant zeros of $\Sigma$ are contained in $\mathds C^-$, the system is said to have the \emph{minimum phase} property.

\subsection{Strong Detectability and Observability}\label{sec:strongdetectability}

This section recalls the basic concepts of strong detectability and strong observability and discusses several important aspects of systems with these properties. 

\begin{definition}[strong detectability \cite{hautus1983strong}]
	System~\eqref{eq:sys:orig} is called
	\begin{enumerate}[label={\roman*)}]
		\item \emph{strongly observable}, if $\y(t)= \bm 0\text{ for all } t\geq 0$ implies $\x(t)=\bm 0$ for all $t\geq 0$, all $\u(t)$ and all $\x(0)=\x_0$;
		\item \emph{strongly detectable}, if $\y(t)= \bm 0$ for all $t\geq 0$ implies $\x(t)\rightarrow \bm 0$ for $t\rightarrow\infty$, all $\u(t)$ and all $\x(0)=\x_0$;		
		\item \emph{strong$^*$ detectable}, if $\y(t)\rightarrow \bm 0$ for $t\rightarrow\infty$ implies $\x(t)\rightarrow \bm 0$ for $t\rightarrow\infty$, all $\u(t)$ and all $\x(0)=\x_0$.
	\end{enumerate}
\end{definition}

It is well known that strong$^*$ detectability is the minimum requirement for the existence of a linear unknown input observer~\cite{hautus1983strong}, whereas strong detectability is the minimum requirement for (at least asymptotically) reconstructing the state with any estimation scheme~\cite{bejarano2009unknown}.
If the system is strongly observable, it is possible to exactly reconstruct the state within finite or fixed time by employing higher-order sliding mode techniques~\cite{fridman2011high-order,bejarano2009unknown,tranninger2019exact}.
The following relations are stated in~\cite{hautus1983strong}. 
\begin{proposition}[strong observability and detectability conditions]
	System $\Sigma$ is
	\begin{enumerate}[label=(\roman*)]
		\item strongly observable, if and only if $\rank \P(\lambda)=n+m$ for all $\lambda\in\mathds C$. 
		\item strongly detectable, if and only if $\rank \P(\lambda)=n+m$ for all $\lambda\in\mathds C$ with $\Re{\lambda}\geq 0$.
		\item strong$^*$ detectable, if and only if it is strongly detectable and additionally 
		\begin{equation}\label{eq:rankcond}
			\rank \begin{bmatrix} \C\B & \D \\
				\D & \bm 0 \end{bmatrix} = \rank \D + \rank\begin{bmatrix} \B \\ \D \end{bmatrix} = \rank\D + m.
		\end{equation}
	\end{enumerate}
\end{proposition}
\begin{remark}
	For the characterization of strong observability and detectability,~\cite{hautus1983strong} introduces a slightly different definition of zeros than in Definition~\ref{def:invariantzero}. 
	Both definitions coincide under the assumption that $\normrank \P(s)=n+m$. 
	In this case, condition \textit{(i)} essentially requires that the system has no invariant zeros. 
	Condition \textit{(ii)} then states that the system is strongly detectable if and only if it is minimum phase.
	Moreover, $\normrank \P(s)=n+m$ if and only if $\normrank \G(s) = m$. 
	A system for which the latter relation holds is also called \emph{left-invertible}. 
	A necessary condition for left-invertability of $\G(s)$ is that $p\geq m$, i.e., that there are at least as many linearly independent measurements as unknown inputs. 
	Consequently, this condition is also necessary for strong detectability.
\end{remark}

Condition~\eqref{eq:rankcond} is the so-called rank-condition and is a basic requirement for the design of a linear unknown input observer without using derivatives of the output signal.
One can see from the above conditions that strong observability implies strong detectability. 
Moreover, strong$^*$ detectability implies strong detectability. 
The converse is not true as shown with an example in~\cite{hautus1983strong}.
An alternative characterization of strong observability and detectability can be given in terms of invariant subspaces:
\begin{definition}[weakly unobservable subspace~\cite{trentelman2012control}] \label{def:weaklyunobsv}\leavevmode
	A point $\x_0\in \mathds R^n$ is called weakly unobservable if there exists an input $\u(t)$ such that the corresponding output satisfies $\y(t)= \bf0$ for $t\geq 0$ and $\x(0)=\x_0$. 
	The set of all weakly unobservable points is denoted by $\mathcal V^*(\Sigma)$ and is called the weakly unobservable subspace of $\Sigma$. 
\end{definition}

\begin{definition}[controllable weakly unobservable subspace~\cite{trentelman2012control,chen2004linear}]\label{def:contrweaklyunobsv} \leavevmode
A point $\x_0\in\mathds R^n$ is called controllable weakly unobservable, if there exists an input signal $\u(t)$ and a $T>0$, such that $\y(t) = \bf 0$ for all $t\in[0,T]$ and $\x(T)=\bf 0$. 
The set of all controllable weakly unobservable points is denoted by $\mathcal R^*(\Sigma)$ and is called the controllable weakly unobservable subspace of $\Sigma$.
\end{definition}
It follows directly from Definitions~\ref{def:weaklyunobsv} and~\ref{def:contrweaklyunobsv} that $\mathcal{R}^*(\Sigma) \subseteq \mathcal{V}^*(\Sigma)$. 
A thorough introduction to the geometric subspace approach for linear systems is given in the book of Trentelman, Stoorvogel and Hautus~\cite{trentelman2012control}.
There, the following two results are presented
\begin{lemma}[geometric conditions for strong observability and strong detectability~{{\cite[Ch. 7]{trentelman2012control}}}]\leavevmode
\begin{enumerate}
	\item[i)] $\Sigma$ is strongly observable if and only if $\mathcal{V}^*(\Sigma) = 0$.
	\item[ii)] $\Sigma$ is strongly detectable if and only if $\mathcal R^*(\Sigma)=0$ and $\Sigma$ is minimum phase.
\end{enumerate}
\end{lemma}

\subsection{Moreno's Arbitrary Order Fixed-Time Differentiator}\label{sec:morenodiff}
Recently, J. A. Moreno proposed a family of finite-/fixed-time convergent differentiators or observers in~\cite{moreno2020arbitrary}.
Fixed-time convergence is often desirable, if no bounds on the initial conditions are known.
In this case, the finite-time convergent observers may take arbitrarily long to converge because the convergence time grows with the initial condition; fixed-time convergence guarantees an upper bound for the convergence time.
Finite-time convergent observers and in particular Levant's robust exact differentiator (RED) are included as special cases within the family of arbitrary order finite-/fixed-time differentiators.
Important results from~\cite{moreno2020arbitrary} are recalled in the following.

The system under consideration is given by the integrator chain
\begin{subequations}
	\begin{align}
		\dot x_1 &= x_2,\quad y = x_1 \\
		\dot x_2 &= x_3, \\
		&\, \vdots \\
		\dot x_n &= u,	
	\end{align}	
\end{subequations}
with the state $\x = \begin{bmatrix}
	x_1 & x_2& \cdots & x_n
\end{bmatrix}^\transp$ and $n$ as the system order. 
The input signal $u$ is assumed to be bounded with $|u(t)| \leq \Delta$ for all $t \geq 0$ and some constant $\Delta\geq 0$. 
The observer based on~\cite{moreno2020arbitrary} can then be stated as
\begin{subequations}
	\begin{align}
		\dot{\hat x}_1 &= \hat x_2 + \kappa_1\Phi_1^n(e_1),\\
		\dot{\hat x}_2 &= \hat x_3 + \kappa_2\Phi_2^n(e_1), \\
		&\,\vdots \\
		\dot{\hat x}_n &= \kappa_n\Phi_n^n(e_1),
	\end{align}
\end{subequations}
with positive parameters $\kappa_i$ and nonlinear output injection terms $\Phi_i^n:\mathds R\mapsto \mathds R$, $i=1,\ldots,n$.
The output injection terms are given by
\begin{equation}\label{eq:diff:moreno:phi:1}
	\Phi_i^n(z) = \left(\phi_i^n\,\circ \,\cdots \,\circ\, \phi_2^n\,\circ\,\phi_1^n\right)(z)
\end{equation}
with the monotonic growing functions
\begin{equation}\label{eq:diff:moreno:phi:2}
	\phi_i^n(z) = \mu \lfloor z \rceil^{\frac{r_{0,i+1}}{r_{0,i}}} + (1-\mu) \lfloor z \rceil^{\frac{r_{\infty,i+1}}{r_{\infty,i}}}
\end{equation}
and a constant parameter\footnote{This is a special choice discussed in the introduction of~\cite{moreno2020arbitrary}.} $0<\mu<1$. 
The powers are completely determined by two parameters $-1\leq d_0 \leq d_\infty \leq \frac{1}{n-1}$ according to the recursive definitions
\begin{subequations}\label{eq:diff:moreno:phi:3}
	\begin{align}
		r_{0,i} &= r_{0,i+1} - d_0 = 1-(n-i)d_0, \\
		r_{\infty,i} &= r_{\infty,i+1}-d_\infty = 1-(n-i) d_\infty,
	\end{align}
\end{subequations}
for $i= 1,\ldots,n+1$.
An insightful discussion of the design idea can be found in the introductory section of~\cite{moreno2020arbitrary}.
The dynamics of the estimation error $\e = \x-\hat \x$ can be stated according to
\begin{subequations}\label{eq:error:FT}
	\begin{align}
		\dot e_1 &= e_2 - \kappa_1 \Phi_1^n(e_1) \\ 
		\dot e_2 &= e_3 - \kappa_2 \Phi_2^n(e_1) \\ 
		& \, \vdots \\
		\dot e_n &= u - \kappa_n\Phi_n^n(e_1).
	\end{align}
\end{subequations}
It follows from~\cite[Theorem 1]{moreno2020arbitrary} that there exist appropriate gains $\kappa_i>0$ for $i =1,\ldots, n$, such that~\eqref{eq:error:FT} is asymptotically stable and converges to zero in fixed time if
\begin{enumerate}[label={\roman*)}]
	\item $-1<d_0<0<d_\infty<\frac{1}{n-1}$ and $\Delta = 0$, or 
	\item $-1=d_0 <0<d_\infty < \frac{1}{n-1}$ and $\Delta \geq 0$.
\end{enumerate}
The error dynamics~\eqref{eq:error:FT} is bi-homogeneous as defined in~\cite{andrieu2008homogeneous}, in the sense that near to the origin it is approximated by a homogeneous system of degree $d_0$ and far from the origin its approximation corresponds to a system with homogeneity degree $d_\infty$, see~\cite{moreno2020arbitrary}.
A sequence of stabilizing gains $\kappa_i$, $i=1,\ldots,n$ can be selected according to~\cite[Proposition 4]{moreno2020arbitrary}.
Note, that for $d_0=-1$, the error dynamics~\eqref{eq:error:FT} has a discontinuous right hand side.
This allows for robustness with respect to unknown inputs $u$ with $\Delta >0$.
For $d_0=d_\infty=d$, the error dynamics become homogeneous and in particular, $d=-1$ yields the error dynamics of Levant's robust exact differentiator~\cite{levant2003high-order}. 
In this case, the functions $\Phi_i^n$ are given by
\begin{equation}\label{eq:obsv:Levant:Phi}
	\Phi_i^n(z) = \lfloor z \rceil^{\frac{n-i}{n}}\quad \text{for } i = 1,\ldots,n
\end{equation}
and the error dynamics~\eqref{eq:error:FT} read as
\begin{subequations}\label{eq:error:SM}
		\begin{align}
		\dot e_1 &= e_2 - \kappa_1\lfloor e_1 \rceil^{\frac{n-1}{n}} \\ 
		\dot e_2 &= e_3 - \kappa_2 \lfloor e_1 \rceil^{\frac{n-2}{n}} \\ 
		& \, \vdots \\
		\dot e_n &= u - \kappa_n\lfloor e_1 \rceil^{0}.\label{eq:error:SM:last}
	\end{align}
\end{subequations}
For $|u|\leq \Delta$ and suitable gains $\kappa_i>0$, the solution of the error dynamics~\eqref{eq:error:SM} converges to zero in finite-time~\cite{levant1998robust,levant2003high-order}.
Well established parameter settings for the differentiator gains up to order $n=6$ are proposed in~\cite[Section 6.7]{shtessel2013sliding}.
A necessary condition for any suitable parameter set is $\kappa_n > \Delta$ and hence $\kappa_n = 1.1\Delta$ is proposed in~\cite{levant2003high-order}. 
In this case, the discontinuous error injection in~\eqref{eq:error:SM:last} is able to dominate the unknown input $u$.

\section{The Special Coordinate Basis}\label{sec:SCB}

The special coordinate basis (SCB) was introduced by Sannuti and Saberi~\cite{sannuti1987special} in order to investigate structural properties of LTI systems.
After transformation to the SCB, the state space is decomposed into four parts $\mathds R^{n} = \mathcal X_\mathrm{a} \oplus \mathcal X_\mathrm{b}\oplus \mathcal X_\mathrm{c} \oplus \mathcal X_\mathrm{d}$ with corresponding state vectors $\x_\mathrm{a}$, $\x_\mathrm{b}$, $\x_\mathrm{c}$ and $\x_\mathrm{d}$.
The related subsystems are denoted by (a), (b), (c) and (d) or $\Sigma_{\mathrm a}$, $\Sigma_{\mathrm b}$, $\Sigma_{\mathrm c}$ and $\Sigma_{\mathrm d}$, respectively.
The SCB reveals explicitly the invariant zeros, which govern the dynamics of subsystem (a).
It also shows the system's invertability structure. 
The system is right invertable if and only if subsystem (b) is non-existent and it is left invertable if and only if subsystem (c) is non-existent.
The properties of the SCB are extensively treated in~\cite[Chapter 5]{chen2004linear} and briefly summarized in Section~\ref{sec:SCB:properties}.

For system~\eqref{eq:sys:orig}, the transformation to the SCB is summarized in
\begin{proposition}[SCB, {\cite[Theorem 5.4.1]{chen2004linear}}]\label{prop:SCB}
	There exist nonsingular state, input and output transformations $\x = \Ts \bar \x$, $\u =\Ti\bar \u$ and $\y=\To\bar \y$ for system~\eqref{eq:sys:orig}, such that the transformed system is given by
	\begin{subequations}\label{eq:scb}
		\begin{align}
			\begin{bmatrix}
				\dot \x_\mathrm{a}\\
				\dot \x_\mathrm{b}\\
				\dot \x_\mathrm{c}\\
				\dot{\x}_\mathrm{d}\\
			\end{bmatrix} &= \left(
			\begin{bmatrix} 
				\Aa & \Hab \Cb & \bm 0 & \Had\Cd \\
				\bm 0   & \Ab      & \bm 0 & \Hbd\Cd \\
				\Bc\Eca & \Hcb\Cb & \Ac & \Hcd\Cd \\
				\Bd\Fda & \Bd\Fdb & \Bd\Fdc & \Ad
			\end{bmatrix} 
			+ \begin{bmatrix}
				\Boa \\
				\Bob\\
				\Boc\\
				\Bod
			\end{bmatrix} \begin{bmatrix}
				\Coa & \Cob & \Coc & \Cod
			\end{bmatrix}%
			\right)
\underbrace{			\begin{bmatrix}
				\x_\mathrm{a}\\
				\x_\mathrm{b}\\
				\x_\mathrm{c}\\
				\x_\mathrm{d}\\
			\end{bmatrix}}_{\bar \x}
			+
			\begin{bmatrix}
				\Boa & \bm 0 & \bm 0\\
				\Bob & \bm 0 & \bm 0\\
				\Boc & \bm 0 & \Bc \\
				\Bod & \Bd & \bm 0
			\end{bmatrix}
			\underbrace{\begin{bmatrix}
				\u_0\\
				\ud\\
				\uc
			\end{bmatrix}}_{\bar \u},\\
			\underbrace{\begin{bmatrix}
				\y_0\\
				\yd \\
				\yb 
			\end{bmatrix}}_{\bar \y} &= 
			\begin{bmatrix}
				\Coa & \Cob & \Coc & \Cod \\
				\bm 0 & \bm 0 & \bm 0 & \Cd \\
				\bm 0 & \Cb &  \bm0 & \bm 0
			\end{bmatrix}
			\begin{bmatrix}
				\x_\mathrm{a}\\
				\x_\mathrm{b}\\
				\x_\mathrm{c}\\
				 \x_\mathrm{d}\\
			\end{bmatrix}+\begin{bmatrix}
				\I_{m_0} &\bm 0& \bm 0\\
				\bm 0 & \bm 0 & \bm 0\\
				\bm 0 & \bm 0 & \bm 0\\
			\end{bmatrix}\begin{bmatrix}
				\u_0\\
				\ud\\
				\uc
			\end{bmatrix}, \quad \begin{array}{cccc}
				\x_\mathrm{a} \in \mathds R^{\na},&  \x_\mathrm{b} \in \mathds R^{\nb}, &\x_\mathrm{c} \in \mathds R^{\nc},&\x_\mathrm{d} \in \mathds R^{\nd},\\
				\u_0\in\mathds R^{\mnull},&\ud \in\mathds R^{\md}, & \uc \in \mathds R^{\mc}, \\
				\y_0 \in\mathds R^{\mnull},& \yb \in\mathds R^{\pb}, &\yd \in \mathds R^{\md}.
			\end{array}
		\end{align}
	\end{subequations}
All matrices are assumed to be of appropriate dimensions. 
The matrices $\Ab$ and $\Cb$ are given according to 
\begin{equation}
	\Ab = \Ab^\star + \Hbb\Cb \quad \text{and} \quad \Cb = \diag{\C_{l_1},\C_{l_2},\ldots,\C_{l_{\pb}}},
\end{equation}
with 
\begin{equation}
	\Ab^\star = \diag{\J_{l_1},\J_{l_2},\ldots,\J_{l_{\pb}}},\quad \C_{l_i} = \begin{bmatrix}
		1 & \bm{0}_{1\times {l_i-1}}
	\end{bmatrix},\quad i=1,\ldots,\pb
\end{equation}
and positive integers $l_1,l_2,\ldots,l_{\pb}$ such that $\sum_{i=1}^{\pb} l_i = \nb$.
Moreover, $\Hbb$ is a constant $\nb\times \pb$ matrix.
The matrices $\Bd$ and $\Cd$ are given by
\begin{equation}\label{eq:BdCd}
	\Bd=\diag{\B_{q_1},\B_{q_2},\ldots,\B_{q_{\md}}}\quad \text{and}\quad \Cd=\diag{\C_{q_1},\C_{q_2},\ldots,\C_{q_{\md}}}
\end{equation}
with positive integers $q_1\geq q_2 \geq \cdots \geq q_{\md}$, $\sum_{i=1}^{\md} q_i = n_d$, 
\begin{equation}
\B_{q_i} = \begin{bmatrix}
	\bm{0}_{(q_i-1)\times 1} \\
	1
\end{bmatrix} \quad \text{and} \quad \C_{q_i} = \begin{bmatrix}
1 & \bm{0}_{1\times ({q_i-1})}
\end{bmatrix}.
\end{equation}
Moreover,
\begin{equation}
	\Ad = \Ad^\star + \Bd \Fdd + \Hdd \Cd,
\end{equation}
where $\Fdd$ and $\Hdd$ are constant matrices of appropriate dimension and
\begin{equation}\label{eq:Adstar}
\Ad^\star = \diag{\J_{q_1},\J_{q_2},\ldots,\J_{q_{\md}}}.\quad 
\end{equation}
\end{proposition}
A numerically stable algorithm for obtaining the involved transformations is proposed in~\cite{chu2002onthenumerical} and implemented in the Linear Systems Toolkit~\cite{liu2005linear}.
For the unknown input observer design, the system is not yet in a suitable form. 
Before the details of the SCB are discussed, a special choice of the transformation related to subsystem (d) is proposed.
The transformation is introduced in
	\begin{theorem}\label{THM:TRANSF_SUBSYS_D}
		For the transformation to an SCB according to Proposition~\ref{prop:SCB}, 
		the nonsingular state transformation $\x = \Ts \bar \x$ can be chosen such that the matrix $\Fdd$ has the following particular structure:
		\begin{equation}\label{eq:Fdd}
			\Fdd = \begin{bmatrix}
				\begin{array}{cccc;{2pt/2pt}cccc;{2pt/2pt}cccc;{2pt/2pt}ccc}
					0 & 0& \cdots & 0 & 0 & 0 &\cdots & 0 &\multicolumn{4}{c;{2pt/2pt}}{\cdots}   & 0&\cdots & 0\\
					0& \beta_{1,2,1} & \cdots & \beta_{1,2,q_1-1} & 0 & 0 &\cdots & 0 & \multicolumn{4}{c;{2pt/2pt}}{\cdots}  & 0& \cdots &0  \\
					0& \beta_{1,3,1} & \cdots & \beta_{1,3,q_1-1} & 0& \beta_{2,3,1} & \cdots & \beta_{2,3,q_2-1} &   \multicolumn{4}{c;{2pt/2pt}}{\cdots}  & 0& \cdots &0  \\ %
					\multicolumn{4}{c;{2pt/2pt}}{\vdots} &  \multicolumn{4}{c;{2pt/2pt}}{\vdots}  & \multicolumn{4}{c;{2pt/2pt}}{\ddots} &  \multicolumn{3}{c}{\vdots} \\
					0& \beta_{1,\md,1} & \cdots & \beta_{1,\md,q_1-1} & \multicolumn{4}{c;{2pt/2pt}}{\cdots} & 0& \beta_{\md-1,\md,1} & \cdots & \beta_{\md-1,\md,q_{\md}-1} & 0&\cdots & 0 
				\end{array}
			\end{bmatrix} = \begin{bmatrix}
				\bm 0^\transp \\
				\bm \beta_2^\transp\\
				\bm \beta_3^\transp \\
				\vdots \\
				\bm \beta_{\md}^\transp
			\end{bmatrix}
		\end{equation}
	\end{theorem}
In order to proof this result, an auxiliary step is required.
Note, that with this special choice, the system is still in an SCB. 
In particular, if the system is already in an SCB according to Proposition~\ref{prop:SCB}, it is sufficient to apply a change of coordinates to Subsystem $\Sigma_d$. 
This is summarized in
\begin{lemma}\label{LE:SCB_SPECIAL}
Assume that System~\eqref{eq:sys:orig} is already in the SCB presented in Proposition~\ref{prop:SCB}.
Then, there exists a regular state transformation matrix $\x = \Ts \bar \x$ with $\Ts=  \diag{\I_{\na},\I_{\nb},\I_{\nc}, \Td}$, such that $\Fdd$ has the structure presented in Theorem~\ref{THM:TRANSF_SUBSYS_D}.
Moreover, it holds that $\Cd \Td =\Cd$ and $\Td^{-1}\Bd =\Bd$.
\end{lemma}
The proof of Lemma~\ref{LE:SCB_SPECIAL} is based on the proof of~\cite[Theorem 3.1]{niederwieser2021high} and presented in the Appendix.
The proof of Theorem~\ref{THM:TRANSF_SUBSYS_D} then follows straightforwardly from Proposition~\ref{prop:SCB} and Lemma~\ref{LE:SCB_SPECIAL}.

\begin{remark}
In both SCBs presented in Proposition~\ref{prop:SCB} and Theorem~\ref{THM:TRANSF_SUBSYS_D}, subsystem $\Sigma_{\mathrm d}$ is formed by $\md$ chains of integrators with length $q_i$, $i=1,\ldots,\md$ up to an output feedback with feedback matrix $\Hdd$, respectively.
The difference is, that for the special form of $\Fdd$ according to~\eqref{eq:Fdd} in Theorem~\ref{THM:TRANSF_SUBSYS_D}, the resulting matrix $\Ad$ is a block lower triangular matrix up to an output feedback.
This will be essential for the construction of the unknown input observers.
\end{remark}

The SCB reveals important structural system properties, which are discussed in the following. 
Further details can be found in~\cite{sannuti1987special,chen2004linear}

\subsection{Properties of the SCB}\label{sec:SCB:properties}
As already mentioned, the state space of~\eqref{eq:scb} is decomposed into $\mathds R^{n} = \mathcal X_\mathrm{a} \oplus \mathcal X_\mathrm{b}\oplus \mathcal X_\mathrm{c} \oplus \mathcal X_\mathrm{d}$ corresponding to the states $\x_\mathrm{a}$, $\x_\mathrm{b}$, $\x_\mathrm{c}$ and $\x_\mathrm{d}$. 

Subsystem (a) is related to the invariant zeros, i.e., the eigenvalues of $\Aa$. 
Subsystems (b) and (c) are related to the invertability properties of the system and subsystem (d) reveals the infinite zero structure.

Subsystem (b) consists of $\pb$ decoupled chains of integrators up to an output feedback.
More specifically, one can split its state space according to
\begin{equation}
	\x_\mathrm{b} = \begin{bmatrix}
		\x_{\mathrm{b},1}^\transp 	&	\x_{\mathrm{b},2}^\transp  & \cdots & 		\x_{\mathrm{b},\pb}^\transp 
	\end{bmatrix}^\transp
\end{equation}
and each subsystem corresponding to $\x_{\mathrm{b},i}$ for $i= 1,\ldots,\pb$ takes the specific form
\begin{subequations}\label{eq:sysb:intchains}
	\begin{align}
		\dot x_{\mathrm{b},i,1} &= x_{\mathrm{b},i,2} + \h^\transp_{\mathrm{bb},i,1}\yb+\h^\transp_{\mathrm{db},i,1}\yd   \\
		\dot x_{\mathrm{b},i,2} &= x_{\mathrm{b},i,3}+ \h^\transp_{\mathrm{bb},i,2}\yb + \h^\transp_{\mathrm{db},i,2}\yd \\
		& \;\vdots  \\ \nonumber
		\dot x_{\mathrm{b},i,l_i} &= \h^\transp_{\mathrm{bb},i,l_i}\yb + \h^\transp_{\mathrm{db},i,l_i} \yd \\
		y_{\mathrm{b},i} &= x_{\mathrm b,i,1},
	\end{align}
\end{subequations}
with appropriate row vectors $\h^\transp_{\mathrm{bb},i,1}$, $\ldots$, $\h^\transp_{\mathrm{bb},i,l_i}$ and $\h^\transp_{\mathrm{bd},i,1}$, $\ldots$, $\h^\transp_{\mathrm{bd},i,l_i}$.

The infinite zero structure represented by subsystem (d) is of great importance for state estimation in the presence of unknown inputs.
The state $\x_\mathrm{d}$ can be decomposed according to
\begin{equation}
	\x_\mathrm{d} = \begin{bmatrix}
		\x_{\mathrm{d},1}^\transp 	&	\x_{\mathrm{d},2}^\transp  & \cdots & 		\x_{\mathrm{d},\md}^\transp 
	\end{bmatrix}^\transp.
\end{equation}
In particular, each subsystem $\x_{\mathrm{d},i}$ for $i= 1,\ldots,\md$ takes the specific form
\begin{subequations}\label{eq:intchains}
\begin{align}
	\dot x_{\mathrm{d},i,1} &= x_{\mathrm{d},i,2} + \h^\transp_{\mathrm{dd},i,1}\yd   \\
	\dot x_{\mathrm{d},i,2} &= x_{\mathrm{d},i,3} + \h^\transp_{\mathrm{dd},i,2}\yd \\
	 & \;\vdots  \\ \nonumber
	 \dot x_{\mathrm{d},i,q_i} &= \f_{\mathrm{da},i}^\transp \xa + \f_{\mathrm{db},i}^\transp  \x_{\mathrm{b}} +\f_{\mathrm{dc},i}^\transp \x_{\mathrm{c}}  +\f_{\mathrm{dd},i}^\transp \x_{\mathrm{d}} + u_{\mathrm d,i}  \\
	 y_{\mathrm{d},i} &= x_{\mathrm d,i,1},
\end{align}
\end{subequations}
with appropriate row vectors $\h^\transp_{\mathrm{dd},i,1}$, $\ldots$, $\h^\transp_{\mathrm{dd},i,q_i}$, $ \f_{\mathrm{da},i}^\transp$, $\f_{\mathrm{db},i}^\transp $, $\f_{\mathrm{dc},i}^\transp$ and $\f_{\mathrm{dd},i}^\transp$.
Each particular subsystem is an integrator chain of length $q_i$ up to an output injection. 
In other words, $q_i$ corresponds to the number of integrations between the input $u_{\mathrm d,i}$ and the output $y_{\mathrm d,i}$.
The list $S_\infty^\star(\Sigma) = \left\{q_1,q_2,\ldots,q_{a}\right\}$ represents the so-called infinite zero structure of the system in the sense that $\Sigma$ has $\md$ infinite zeros of order $q_1$, $q_2$, $\ldots$, $q_{\md}$~\cite{chen2004linear}.
It should be remarked, that $S_\infty^\star(\Sigma)$ corresponds to the list $\I_4$ of Morse's structural invariant indices as defined in~\cite{morse1973structural,chen2004linear}.
The following lemma summarizes important properties of the SCB, that will be required for the proposed observer design.

\begin{lemma}[SCB properties{{~\cite[Sec. 5.4]{chen2004linear}},~\cite{sannuti1987special}}]\leavevmode
	\begin{enumerate}[label={(p.\arabic*)}]
	 \item $\normrank \G(s)= \mnull+\md$.
	 \item The invariant zeros of $\Sigma$ are the eigenvalues of $\Aa$. \label{it:prop:zeros}
	 \item The pair $(\Ab,\Cb)$ is observable.
	 \item The triple $(\Ad,\Bd,\Cd)$ is strongly observable.
	 \item System $\Sigma$ is left invertible if and only if $\nc = 0$. In this case, $\x_\mathrm{c}$ and $\uc$ are nonexistent.
	 \item $\mathcal{X}_\mathrm{a} \bigoplus \mathcal{X}_\mathrm{c}=\mathcal{V}^*(\Sigma)$\label{it:prop:V}
	 \item $\mathcal{X}_\mathrm{c}=\mathcal R^*(\Sigma)$. \label{it:prop:R}
	\end{enumerate}
\end{lemma}

The following implications follow directly from the properties of the SCB.

\begin{lemma}[strong detectability in SCB]\label{le:strongdetectabilitySCB}
	System~\eqref{eq:sys:orig} is 
	\begin{enumerate}[label={\roman*)}]
		\item strongly observable, if and only if for its SCB~\eqref{eq:scb} it holds that $n_a = n_c =0$.
		\item strongly detectable, if and only if for its SCB~\eqref{eq:scb} it holds that $n_c=0$ and $\Aa$ is a Hurwitz matrix.
		\item strong$^*$ detectable, if and only if it is strongly detectable and for its SCB~\eqref{eq:scb} it holds that $q_i = 1$ for all $i=1,\ldots,\md$.
	\end{enumerate}
\end{lemma}
\begin{proof}
Item i) follows directly from SCB property \ref{it:prop:V}. 
Item ii) is due to properties \ref{it:prop:zeros} and \ref{it:prop:V} and it remains to prove  item iii). 
It is assumed that system~\eqref{eq:scb} is strongly detectable, and hence $\nc = 0$. 
The rank condition~\eqref{eq:rankcond} is equivalent to the condition
\begin{equation}\label{eq:rank_v2}
	\rank \begin{bmatrix}
		\K\C\B \\ \D
	\end{bmatrix} = \rank \begin{bmatrix}
	\B\\ \D
\end{bmatrix}, 
\end{equation}
where $\K$ is a $(p-\mnull)\times p$ full row rank matrix satisfying $\K\D=\bm 0$, see~\cite[Proposition 4]{valcher1999state}. 
For system~\eqref{eq:scb} in the SCB, one can choose
\begin{equation}
	\K = \begin{bmatrix}
		\bm 0_{\md\times \mnull} & \I_{\md} & \bm 0 \\
		\bm 0 & \bm 0 & \I_{\pb} 
		\end{bmatrix}
\end{equation}
and hence condition~\eqref{eq:rank_v2} results in
\begin{equation}
	\rank \begin{bNiceMatrix}
		\Cd\Bod & \Cd\Bd \\
		\Cb\Bob & \bm 0 \\ \hdottedline
		\I_{\mnull} & \bm 0\\
		\bm 0 & \bm 0 \\ 
		\bm 0 & \bm 0
	\end{bNiceMatrix}
 = \mnull+\md.
\end{equation}
This condition is equivalent to $\rank \Cd\Bd = \md$.
Due to the structure of $\Cd$ and $\Bd$, the latter condition can be fulfilled, if and only if $q_i=1$ for all $i=1,\ldots,\md$, i.e., if the length of each integrator chain in subsystem (d) is equal to 1.
\end{proof}

\section{Observer Design}\label{sec:observerdesign}
Because strong detectability is the minimum requirement for the existence of any estimation scheme in the presence of unknown inputs~\cite{bejarano2009unknown}, this property is assumed in the following.
Hence, $\nc=0$, which yields the new system representation:
	\begin{subequations}\label{eq:scb:SD}
	\begin{align}
		\begin{bmatrix}
			\dot \x_\mathrm{a}\\
			\dot \x_\mathrm{b}\\
			\dot \x_\mathrm{d}\\
		\end{bmatrix} &= \left(
		\begin{bmatrix} 
			\Aa & \Hab \Cb &  \Had\Cd \\
			\bm 0   & \Ab      & \Hbd\Cd \\
			\Bd\Fda & \Bd\Fdb &  \Ad 
		\end{bmatrix} 
		+ \begin{bmatrix}
			\Boa \\
			\Bob\\
			\Bod
		\end{bmatrix} \begin{bmatrix}
			\Coa & \Cob  & \Cod
		\end{bmatrix}%
		\right)
		\begin{bmatrix}
			\x_\mathrm{a}\\
			\x_\mathrm{b}\\
			\x_\mathrm{d}\\
		\end{bmatrix}
		+
		\begin{bmatrix}
			\Boa & \bm 0 \\
			\Bob & \bm 0 \\
			\Bod & \Bd  
		\end{bmatrix}
		\begin{bmatrix}
			\u_0\\
			\ud\\
		\end{bmatrix},\label{eq:scb:SD:states}\\
		\begin{bmatrix}
			\y_0\\
			\yd\\
			\yb 
		\end{bmatrix} &= 
		\begin{bmatrix}
			\Coa & \Cob & \Cod \\
			\bm 0 & \bm 0  & \Cd \\
			\bm 0 & \Cb &   \bm 0
		\end{bmatrix}
		\begin{bmatrix}
			\x_\mathrm{a}\\
			\x_\mathrm{b}\\
			\x_\mathrm{d}\\
		\end{bmatrix}+\begin{bmatrix}
			\I_{m_0} &\bm 0\\
			\bm 0 & \bm 0 \\
			\bm 0 & \bm 0 \\
		\end{bmatrix}\begin{bmatrix}
			\u_0\\
			\ud\\
		\end{bmatrix}, \quad \begin{array}{cccc}
			\x_\mathrm{a} \in \mathds R^{\na},&  \x_\mathrm{b} \in \mathds R^{\nb},&\x_\mathrm{d} \in \mathds R^{\nd},\\
			\u_0\in\mathds R^{\mnull},&\ud \in\mathds R^{\md}, \\
			\y_0 \in\mathds R^{\mnull},& \yb \in\mathds R^{\pb}, &\yd \in \mathds R^{\md}.\label{eq:scb:SD:out}
		\end{array}
	\end{align}
\end{subequations}
The overall system order is $n =\na+\nb+\nd$.
By using the output relations~\eqref{eq:scb:SD:out} in~\eqref{eq:scb:SD:states}, system~\eqref{eq:scb:SD} reduces to the three subsystems $\Sigma_{\mathrm a}$, $\Sigma_{\mathrm b}$ and $\Sigma_{\mathrm d}$.
The first subsystem
\begin{equation}\label{eq:sys:a}
	\Sigma_{\mathrm a}:\; \begin{aligned}
		\dot \x_\mathrm{a} &= \Aa\x_\mathrm{a} + \Hab\yb +\Had \yd +\Boa \y_0 
	\end{aligned}
\end{equation}
has no unknown input and no output. 
It is, however, influenced by the outputs of the other subsystems and the output $\y_0$ given by
\begin{equation}\label{eq:sys:y0}
\y_0 = \Coa \x_\mathrm{a} + \Cob\x_\mathrm{b} + \Cod\x_\mathrm{d} + \I_{\mnull} \u_0
\end{equation}
Due to the assumption of strong detectability, the matrix $\Aa$ is a Hurwitz matrix as shown in Lemma~\ref{le:strongdetectabilitySCB}, item ii).
Subsystem $\Sigma_{\mathrm b}$ is given by
\begin{equation}\label{eq:sys:b}
	\Sigma_{\mathrm b}:\;\left\{\begin{aligned}
		\dot \x_\mathrm{b} &= \Ab\x_\mathrm{b} + \Hbd\yd + \Bob \y_0 \\
		\yb &= \Cb\x_\mathrm{b}.
	\end{aligned}\right.
\end{equation}
Again, this subsystem is not directly influenced by the unknown input, but by the outputs $\yd$ and $\y_0$.
The last subsystem 
\begin{equation}\label{eq:sys:d}
	\Sigma_{\mathrm d}:\;\left\{\begin{aligned}
		\dot \x_\mathrm{d} &= \Ad\x_\mathrm{d} + \Bd(\Fda\x_\mathrm{a}+\Fdb\x_\mathrm{b}+\ud) + \Bod\y_0\\		
		\yd &= \Cd\x_\mathrm{d}
	\end{aligned}	\right.
\end{equation}
is influenced by the unknown input $\ud$ and the states $\x_{\mathrm a}$ and $\x_{\mathrm b}$ of the subsystems $\Sigma_{\mathrm a}$ and $\Sigma_{\mathrm b}$, respectively.
With this representation, it is possible to design separate observers for subsystems $\Sigma_{\mathrm a}$, $\Sigma_{\mathrm b}$ and $\Sigma_{\mathrm d}$.

\begin{figure}
	\centering
	\includegraphics[width=0.6\linewidth]{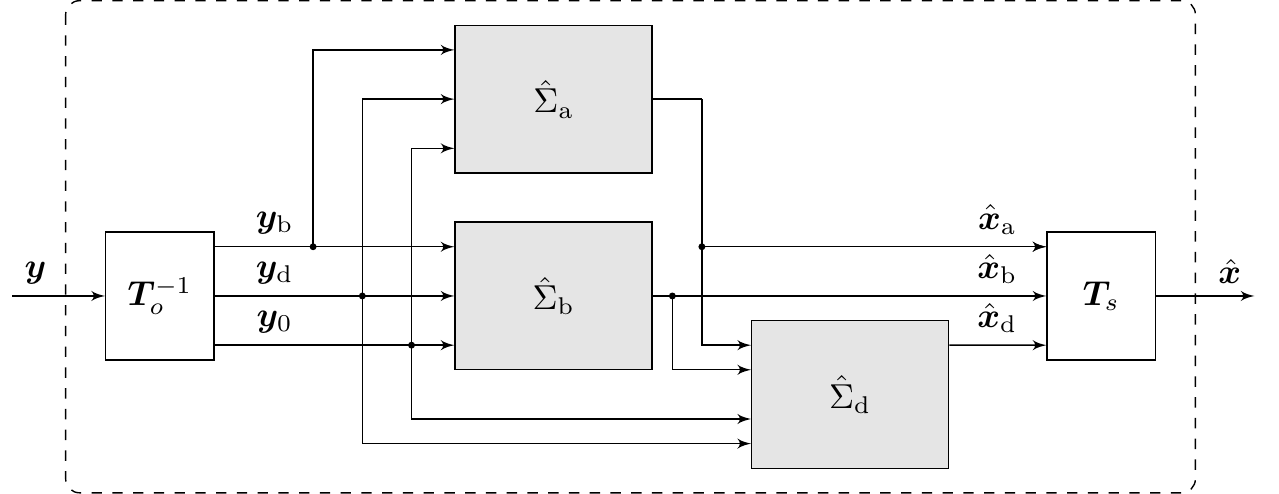}
	\caption{Structure of the proposed observer in the original coordinates.}
	\label{fig:structure}
\end{figure}

For subsystem $\Sigma_{\mathrm a}$, a trivial observer of the form 
\begin{equation}\label{eq:obsv:a}
\hat\Sigma_{\mathrm a}:\;	\dot{\hat\x}_\mathrm{a} = \Aa {\hat\x}_\mathrm{a} + \Hab \yb+\Had \yd+\Boa\y_0
\end{equation}
is chosen and it directly follows that the dynamics of the estimation error $\e_\mathrm{a} = \x_\mathrm{a}-\hat\x_\mathrm{a} $ given by 
\begin{equation}
	\dot\e_\mathrm{a} = \Aa \e_\mathrm{a}
\end{equation}
are asymptotically stable.
The proposed generic observer structure in the original (not SCB) coordinates is depicted in Fig.~\ref{fig:structure}.
The observers $\hat\Sigma_{\mathrm b}$ and $\hat\Sigma_{\mathrm d}$ for subsystems $\Sigma_{\mathrm b}$ and $\Sigma_{\mathrm d}$, respectively, are discussed in the subsequent sections.
For strong$^*$ detectable systems, a linear observer design is presented in Section~\ref{sec:linUIO}.
If the system is merely strongly detectable, derivatives are required in order to estimate the states of $\Sigma_d$.
To that end, a sliding mode observer is proposed in Section~\ref{sec:asymptotic_SMO}.
If the system is strongly observable, it is possible to design fixed-time convergent observers as proposed in Section~\ref{sec:finite/fixed}.
The proposed design procedure is summarized in Section~\ref{sec:designsummary}.

\subsection{Asymptotic Linear Observer Design for Strong* Detectable Systems}\label{sec:linUIO}

It is well known, that a linear unknown input observer exists if and only if the system is strong$^*$ detectable~\cite{hautus1983strong}.
For subsystem (b), a Luenberger-type observer\footnote{For finite- or fixed-time convergence, a nonlinear observer is proposed in Section~\ref{sec:finite/fixed} instead.} of the form
\begin{equation}\label{eq:obsv:b}
	\hat\Sigma_{\mathrm b}:\;	\dot{\hat\x}_\mathrm{b} = \Ab \hat\x_\mathrm{b} + \Hbd\yd + \Bob\y_0 +\Lb(\yb-\Cb\hat\x_\mathrm{b})
\end{equation}
with the output injection matrix $\Lb \in \mathds{R}^{\nb\times \pb}$ is proposed.
The corresponding estimation error dynamics is governed by
\begin{equation}
	\dot\e_\mathrm{b} = (\Ab-\Lb\Cb)\e_\mathrm{b}.
\end{equation}
It follows from the SCB that $(\Ab,\Cb)$ is observable, and hence $\Lb$ can be chosen such that $(\Ab-\Lb\Cb)$ is a Hurwitz matrix with arbitrary eigenvalues.
Consequently, the resulting error dynamics is asymptotically stable and the convergence speed can be assigned arbitrarily.
A proper design of $\Lb$ is required throughout the rest of the paper and summarized in
\begin{assumption}\label{as:Lb}
	The output injection matrix $\Lb$ is chosen, such that $(\Ab-\Lb\Cb)$ is a Hurwitz matrix.
\end{assumption}

Because of the absence of subsystem (c), it remains to design an observer for subsystem (d).
According to Lemma~\ref{le:strongdetectabilitySCB}, $q_i = 1$ for $i=1,\ldots,\md$ in this case.
Together with~\eqref{eq:intchains}, one immediately concludes that $\Bd = \Cd = \I_{\md}$ and $\yd = \x_\mathrm{d}$.
Hence, one may utilize a reduced order observer, i.e., 
\begin{equation}\label{eq:obsv:dred}
\hat\Sigma_{\mathrm d}:\;	\hat \x_\mathrm{d} = \yd.
\end{equation}
This leads to the following corollary
\begin{corollary}\label{co:reduio}
If system~\eqref{eq:scb} is strong$^*$ detectable and Assumption~\ref{as:Lb} holds, i.e., $\Lb$ is designed such that $(\Ab-\Lb\Cb)$ is a Hurwitz matrix, then,~\eqref{eq:obsv:a},~\eqref{eq:obsv:b},~\eqref{eq:obsv:dred} is an asymptotic unknown input observer for~\eqref{eq:scb}.
\end{corollary}

\begin{remark} If one is interested in a full order observer design, it suffices to design a classical linear unknown input observer for the strongly observable subsystem (d), i.e., the triple $(\Ad,\Bd,\Cd)$.
This is possible, because $\Bd = \Cd =\I_{\md}$ and the rank condition $\rank \Cd\Bd = \rank\Bd = \md$ is trivially fulfilled.
Following the design procedure proposed in~\cite{chen1999robust} (see also the algorithm in~\cite[Table 3.1]{chen1999robust}), one can verify that 
\begin{subequations}
 \begin{align}
 	\dot\z_\mathrm{d} &= \F \z_\mathrm{d}, \quad \z_\mathrm{d}(0) = \z_{\mathrm{d},0}\in\mathds R^{\nd}, \\
 	\hat \x_\mathrm{d} &= \z_\mathrm{d} + \yd,
 \end{align}
\end{subequations}
with an arbitrary Hurwitz matrix $\F$ is the resulting full order unknown input observer for subsystem (d). %
For $\z_{\mathrm{d},0} = \bm 0$, the observer reduces to~\eqref{eq:obsv:dred}, which suggests that there is no benefit in designing a full-order unknown input observer in comparison with the reduced order observer proposed in Corollary~\ref{co:reduio}.
This is due to the direct feed-through of the output $\yd$ to the estimate $\hat\x_\mathrm{d}$ which, contrary to the case without unknown inputs, doesn't allow to mitigate effects from, e.g., measurement noise acting on this output by using a dynamic observer.
\end{remark}
If the rank condition is not fulfilled, derivatives of the output $\yd$ are required in order to reconstruct the state $\x_\mathrm{d}$. 
To that end, a higher order sliding mode observer design is proposed in the following.
\subsection{Asymptotic Sliding Mode Observer Design for Strongly Detectable Systems}\label{sec:asymptotic_SMO}
If system~\eqref{eq:scb} is not strong$^*$ detectable but merely strongly detectable, it is still possible to reconstruct the state $\x_\mathrm{d}$ in finite time.
In order to achieve this, derivatives of the output $\yd$ are required. 
The goal of this section is to design an observer, which keeps the number of required derivatives at an absolute minimum. 
This can be achieved by employing sliding mode techniques. 
Therefore, a component-wise bound on the unknown input $\ud$ is required. 
Note that the bounds for $\u$ in the original coordinates~\eqref{eq:uibound} are not necessarily symmetric. 
Let the lower and upper bounds for $\u$ be given by $\u_{\min}$ and $\u_{\max}$, respectively.
In order to obtain the bounds in the SCB, it is possible to compute an offset and the remaining symmetric part, i.e.,
\begin{equation}
	\u_\mathrm{o} = \frac{1}{2}\left(\u_{\max}+\u_{\min}\right) \quad \text{and} \quad \u_\mathrm{s} = \frac{1}{2} \left(\u_{\max}-\u_{\min} \right),
\end{equation}
where one can verify that $\u_{\min}=\u_\mathrm{o}-\u_\mathrm{s}$ and $\u_\mathrm{\max} = \u_\mathrm{o}+\u_\mathrm{s}$.
The input transformation $\bar \u = \Ti^{-1}\u=\G_\mathrm{i}\u$ can be partitioned according to
\begin{equation}
	\begin{bmatrix}
		\u_0\\
		\ud 
	\end{bmatrix} = \begin{bmatrix}
	\G_0 \\
	\G_\mathrm{d}
\end{bmatrix} \u,
\end{equation}
with $\G_\mathrm{d}$ as an $\md\times m$ matrix.
This allows to derive a tight upper bound on the unknown input $\ud$ in the SCB according to
\begin{equation}\label{eq:UIbounds}
	\left|u_{\mathrm d,i} \right| \leq \Delta_{\mathrm d,i} =  \sum_{j=1}^{m} \left|g_{\mathrm d,i,j}\right|u_{\mathrm s,j} + \left|\sum_{j=1}^m g_{\mathrm d,i,j} u_{\mathrm o,j}\right|,
\end{equation}
where $\G_\mathrm{d}=[g_{\mathrm d,i,j}]$.

With $\Ad = \Ad^\star + \Bd\Fdd + \Hdd\Cd$, system $\Sigma_{\mathrm d}$ can be re-written according to
\begin{subequations}
\begin{align}
	\dot \x_\mathrm{d} &= (\Ad^\star + \Bd\Fdd)\x_\mathrm{d} + \Bd(\Fda\x_\mathrm{a} + \Fdb\x_{\mathrm{b}}+\ud) + \Bod \y_0 + \Hdd\yd \\
	\yd &= \Cd \x_\mathrm{d} = \begin{bmatrix} x_{\mathrm d,1,1} & x_{\mathrm d,2,1} & \cdots & x_{\mathrm d,\md,1}\end{bmatrix}^\transp.
\end{align}	
\end{subequations}
The proposed observer for $\Sigma_{\mathrm d}$ is given by
\begin{align}\label{eq:obs:d:SM}
	\hat\Sigma_{\mathrm d}: \quad \dot{\hat \x}_{\mathrm{d}} &= (\Ad^\star + \Bd\Fdd) \hat\x_{\mathrm{d}} +\Bd(\Fda \hat\x_\mathrm{a}+\Fdb \hat{\x}_\mathrm{b}) + \Bod\y_0 + \Hdd \yd + \ld(\e_{\mathrm{d,y}}),
\end{align}	
where
\begin{equation}
	\e_{\mathrm{d,y}} = \yd - \hat\y_\mathrm{d} = \begin{bmatrix}
		e_{\mathrm d,w_1} & e_{\mathrm d,w_2} & \cdots & e_{\mathrm d,w_{\md}}
	\end{bmatrix}^\transp
\end{equation}
is the output error with $w_{i} = \sum_{j=1}^{i-1} q_j + 1$ and $i=1,2,\ldots,\md$.
Moreover, $\ld:\mathds R^{\md} \mapsto\mathds R^{\nd}$ is the nonlinear output injection with
\begin{equation}\label{eq:obsv:SM:injection}
\ld(\e_{\mathrm{d,y}}) = \begin{bNiceArray}{cccc:c:ccc} \kappa_{1,1} \Phi_1^{q_1}(e_{\mathrm d,w_1}) & \cdots & \kappa_{1,q_1-1} \Phi_{q_1-1}^{q_1}(e_{\mathrm d,w_1})&\kappa_{1,q_1} \Phi_{q_1}^{q_1}(e_{\mathrm d,1}) & \cdots & \kappa_{\md,1} \Phi_{1}^{q_{\md}}(e_{\mathrm d,w_{\md}})  & \cdots &\kappa_{\md,q_{\md}} \Phi_{q_{\md}}^{q_{\md}}(e_{\mathrm d,w_{\md}}) 
\end{bNiceArray}^\transp
\end{equation}
with positive parameters $\kappa_{i,j}$, $i=1,\ldots,\md$, $j=0,\ldots,q_{i}-1$.
The nonlinear functions $\Phi_j^{q_i}$ with $i=1,\ldots, \md$ and $j=1,\ldots,q_i$ are designed following the higher order sliding mode approach given by~\eqref{eq:obsv:Levant:Phi}.

\begin{theorem}\label{THM:obs:sm}
	Assume that system~\eqref{eq:scb} with $\Fdd$ as in~\eqref{eq:Fdd} is strongly detectable and Assumption~\ref{as:Lb} holds.
	Then, there exist sufficiently large gains $\kappa_{i,j}>0$ in~\eqref{eq:obsv:SM:injection},
	such that~\eqref{eq:obs:d:SM} is a finite-time unknown input observer for~\eqref{eq:sys:d}, i.e., for every initial condition $\e_{\mathrm d,0}= \x_{\mathrm d,0}-\hat\x_{\mathrm d,0}$, there exists a finite time $T_f$ such that $\e_\mathrm{d}(t) = \x_{\mathrm{d}}(t)-\hat\x_\mathrm{d}(t) = \bm 0$ for all $t\geq T_f$. 
\end{theorem}
\begin{proof}
Let the rows of the matrices $\Fda$ and $\Fdb$ be denoted by
\begin{equation}
	\Fda = \begin{bmatrix}
		\f_{\mathrm{da},1}^\transp \\
		\f_{\mathrm{da},2}^\transp \\
		\vdots\\
		\f_{\mathrm{da},\md}^\transp \\
	\end{bmatrix} \quad\text{and} \quad 	\Fdb = \begin{bmatrix}
	\f_{\mathrm{db},1}^\transp \\
	\f_{\mathrm{db},2}^\transp \\
	\vdots\\
	\f_{\mathrm{db},\md}^\transp \\
\end{bmatrix},
\end{equation}
respectively.
It follows from Theorem~\ref{THM:TRANSF_SUBSYS_D} and the structure of $\Ad$, that the dynamics of the estimation error
\begin{equation}
	\e_\mathrm{d} = \x_\mathrm{d} - \hat \x_{\mathrm{d}} = \begin{bmatrix}
		e_{\mathrm d,1} & e_{\mathrm d,2} & \cdots & e_{\mathrm d,\nd}
	\end{bmatrix}^\transp
\end{equation}
are governed by
\begin{equation}
	\begin{aligned}
	\Sigma_{\mathrm e,\mathrm{d},1}:& \left\{\begin{aligned} %
		\dot e_{\mathrm d,1} &= e_{\mathrm d,2} - \kappa_{1,1}\lfloor e_{\mathrm d,w_1} \rceil^{\frac{q_1-1}{q_1}} \\
		&\,\vdots \\
		\dot e_{\mathrm{d,q_1-1}} &= e_{\mathrm d, q_1} - \kappa_{1,q_1-1}\lfloor e_{\mathrm d,w_1} \rceil^{\frac{1}{q_1}} \\
		\dot e_{\mathrm d,q_1} &= -\kappa_{1,q_1}\lfloor e_{\mathrm d,1} \rceil^{0} + u_{\mathrm d,1} + \f_{\mathrm{da},1}^\transp \e_{\mathrm a} + \f_{\mathrm{db,1}}^\transp \e_{\mathrm b}
\end{aligned}	\right. \\
\Sigma_{\mathrm e,\mathrm{d},2}:& \left\{\begin{aligned}  %
\dot e_{\mathrm d,w_2} &= e_{\mathrm d,w_2+1} - \kappa_{2,1} \lfloor e_{\mathrm d,w_2} \rceil^{\frac{q_2-1}{q_2}}\\
&\,\vdots \\
\dot e_{\mathrm d,w_2+q_2-2} &= e_{\mathrm d,w_2+q_2-1} - \kappa_{2,q_2-1} \lfloor e_{w_2} \rceil^{\frac{1}{q_2}} \\
\dot e_{\mathrm d,w_2+q_2-1} &=  - \kappa_{2,q_2} \lfloor e_{\mathrm d,w_2} \rceil^0 + \bm \beta_2^\transp \e_\mathrm{d}+ u_{d,2} + \f_{\mathrm{da},2}^\transp \e_{\mathrm a} + \f_{\mathrm{db,2}}^\transp \e_{\mathrm b}
\end{aligned}  \right. \\
 &  \vdots \\
\Sigma_{\mathrm e,\mathrm{d},\md}:& \left\{\begin{aligned} 
	\dot e_{\mathrm d,w_{\md}} &= e_{\mathrm d,w_{\md+1}} - \kappa_{\md,1} \lfloor e_{\mathrm d,w_{\md}} \rceil^{\frac{q_{\md}-1}{q_{\md}}}\\
	&\,\vdots \\
	\dot e_{\mathrm d,\nd-1} &= e_{\mathrm d,\nd} - \kappa_{\md,1} \lfloor e_{\mathrm d,w_{\md}}\rceil^{\frac{1}{q_{\md}}} \\
		\dot e_{\mathrm d,\nd} &= - \kappa_{\md,q_{\md}} \lfloor  e_{\mathrm d,w_{\md}} \rceil ^0+\bm \beta_{\md}^\transp \e_\mathrm{d} + u_{d,\md} + \f_{\mathrm{da},\md}^\transp \e_{\mathrm a} + \f_{\mathrm{db,\md}}^\transp \e_{\mathrm b}
\end{aligned}  \right.
\end{aligned}
\end{equation}
The dynamics of each subsystem $\Sigma_{\mathrm e,\mathrm d,i}$, $i=1,\ldots,\md$ coincides with that of the robust exact differentiator~\eqref{eq:error:SM}.
Note that the errors $\e_\mathrm{a}$ and $\e_\mathrm{b}$ decay exponentially. 
Hence, for every $\epsilon>0$ there exists a finite time $T_\epsilon$, such that $|\f_{\mathrm{da},i}^\transp\e_\mathrm{a}(t)+\f_{\mathrm{db},i}^\transp\e_\mathrm{b}(t)|< \epsilon$ for all $i=1,\ldots,\md$ and $t\geq T_\epsilon$.

Now, consider the first subsystem $\Sigma_{\mathrm{e,d,}1}$. 
Define a new unknown input $\tilde{u}_{\mathrm{d,1}}$ according to
\begin{equation}
	\tilde{u}_{\mathrm d,1} = u_{\mathrm d,1}+ \f_{\mathrm{da},1}^\transp \e_{\mathrm a} + \f_{\mathrm{db,1}}^\transp \e_{\mathrm b}.
\end{equation}
For any (arbitrarily small) $\epsilon >0$, there exists a finite time $T_{\epsilon,1}$ such that this input is bounded by $|\tilde{u}_{\mathrm d,1}| \leq \Delta_{\mathrm d,1} + \epsilon$.
Hence, there exist sufficiently large gains $\kappa_{1,1},\ldots,\kappa_{1,q_1}$ such that the states in $\Sigma_{\mathrm e, \mathrm d, 1}$ are exactly zero after a finite transient time.

The unknown input acting on $\Sigma_{\mathrm{e,d,}2}$ is given by
\begin{equation}
	\tilde{u}_{\mathrm d,2} = \bm \beta_2^\transp \e_\mathrm{d}+ u_{d,2} + \f_{\mathrm{da},2}^\transp \e_{\mathrm a} + \f_{\mathrm{db,2}}^\transp \e_{\mathrm b}.
\end{equation}
It follows from the structure of $\Fdd$ given in~\eqref{eq:Fdd}, that the first part $\bm \beta_2^\transp \e_\mathrm{d}$ only depends on the states in $\Sigma_{\mathrm{e,d,}1}$ and hence this term vanishes after a finite transient time.
Consequently, there exists a finite time $T_{\epsilon,2}$ such that the input is bounded by $|\tilde{u}_{\mathrm d,2}|\leq \Delta_{\mathrm d,2} + \epsilon$ and the states of this subsystem converge to zero in finite time for sufficiently large gains $\kappa_{2,1},\ldots,\kappa_{2,q_2}$.
The rest of the proof follows analogously by induction.
\end{proof}

\begin{corollary}
Suppose that~\eqref{eq:scb} is strongly detectable, that Assumption~\ref{as:Lb} holds and that $\hat\Sigma_\mathrm{d}$ is designed according to~\eqref{eq:obs:d:SM} and Theorem~\ref{THM:obs:sm}.
Then,~\eqref{eq:obsv:a},~\eqref{eq:obsv:b} and~\eqref{eq:obs:d:SM} is an asymptotic unknown input observer for~\eqref{eq:scb}.
\end{corollary}

\begin{remark}\label{remark:1smo}
	If $q_i=1$ for some $\Sigma_{\mathrm e,\mathrm d,i}$, the corresponding error dynamics reduces to a first order sliding mode dynamics.
	Hence, if the system is strong$^*$ detectable, i.e., $q_i= 1$ for all $i=1,\ldots,\md$, this design procedure results in a first order sliding mode observer (1-SMO).
\end{remark}

The proposed unknown input observer design can be seen as a generalization of~\cite{niederwieser2021high}, which considers only the strongly observable case.
It has some advantages compared to already existing higher order sliding mode observers, which are discussed in the following.
Compared to~\cite{niederwieser2021high}, its construction builds upon the special coordinate basis.
This form is well studied in the literature and there are numerically reliable algorithms to obtain the transformation~\cite{chu2002onthenumerical}.
Moreover, the direct feed-through case is explicity included within the proposed framework.
In contrast to~\cite{bejarano2009unknown,bejarano2010high-order,ferreira2011robust,fridman2011high-order,tranninger2019exact}, it does not require the design of an additional ``stabilizing'' Luenberger observer, see~\cite{niederwieser2021high}.
This reduces the design complexity and simplifies the tuning procedure.
Moreover, the infinite zeros structure $S_\infty^\star(\Sigma) = \left\{q_1,q_2,\ldots,q_{a}\right\}$ or equivalently the list $\I_4$ of Morse's structural invariant indices, see Section~\ref{sec:SCB}, represents the number of required signal derivatives. 
This is the minimum number of derivative required for the reconstruction of the states.
In the proposed design, the derivatives of the output error signal $\e_{\mathrm{d,y}}$ are obtained component-wise with possibly distinct differentiator orders. 
In contrast to~\cite{bejarano2010high-order}, which requires the least number of vector-valued derivatives, our observer architecture typically requires less derivatives if the lengths of the integrator chains in subsystem (d) are different.
In contrast to works like~\cite{bejarano2009unknown} or~\cite{fridman2011high-order}, the design is based on the SCB representation. 
This allows to use sophisticated and numerically reliable algorithms to transform the system into the desired form~\cite{chen2004linear,chu2002onthenumerical}.
Moreover, the proposed design could be easily extended in the sense that if some $q_i=1$, i.e., the corresponding state can be directly measured, it is possible to use this measurement in the spirit of the reduced order unknown input observer design in Section~\ref{sec:linUIO}.
In practice, this reduces chattering effects.

For strongly observable systems, fixed-time estimation can be achieved by utilizing Moreno's fixed-time differentiator.
The corresponding design procedure is presented in the following section.

\subsection{Fixed-Time Convergent Observer Design for Strongly Observable Systems}\label{sec:finite/fixed}

If system~\eqref{eq:sys:orig} is strongly observable, its SCB reveals that subsystems $\Sigma_{\mathrm a}$ and $\Sigma_{\mathrm c}$ are absent, see Lemma~\ref{le:strongdetectabilitySCB}.
The system in the SCB coordinates is then given by the subsystems 
\begin{subequations}\label{eq:sys_SO}
\begin{equation}\label{eq:sys_SO:b}
	\Sigma_{\mathrm b}:\;\left\{\begin{aligned}
		\dot \x_\mathrm{b} &= \Ab^\star\x_\mathrm{b} + \Hbb\yb +\Hbd\yd + \Bob \y_0\phantom{1} \\
		\yb &= \Cb\x_\mathrm{b}
	\end{aligned}\right.
\end{equation}
and
\begin{equation}\label{eq:sys_SO:d}
	\Sigma_{\mathrm d}:\;\left\{\begin{aligned}
		\dot \x_\mathrm{d} &= \Ad\x_\mathrm{d} + \Bd(\Fdb\x_\mathrm{b}+\ud) + \Bod\y_0,\\		
		\yd &= \Cd\x_\mathrm{d},
	\end{aligned}	\right.
\end{equation}
with
\begin{equation}\label{eq:sys_SO:y0}
	\y_0 = \Cob\x_\mathrm{b} + \Cod\x_\mathrm{d} + \I_{\mnull} \u_0.
\end{equation}
\end{subequations}
The goal is to design a fixed-time observer.
To this end, and, because subsystem $\Sigma_{\mathrm b}$ is not influenced by the unknown input, the following continuous bi-homogeneous observer (CBHO) is proposed for subsystem $\Sigma_{\mathrm b}$:
\begin{equation}\label{eq:obsv:FT:b}
	\hat\Sigma_{\mathrm b}:\;	\dot{\hat\x}_\mathrm{b} = \Ab^\star \hat\x_\mathrm{b} + \Hbb\yb +\Hbd\yd + \Bob\y_0 +\lb(\e_{\mathrm{b,y}}),
	\end{equation}
where 
\begin{equation}
	\e_{\mathrm{b,y}} = \yb- \hat\y_{\mathrm b} = \begin{bmatrix}
		e_{\mathrm b,r_1} & e_{\mathrm b,r_2} & \cdots & e_{\mathrm b,r_{\pb}}
	\end{bmatrix}^\transp
\end{equation}
with $r_{i} = \sum_{j=1}^{i-1}l_j + 1$ and $i=1,2,\ldots,\pb$.
The nonlinear output injection $\lb:\mathds{R}^\pb \mapsto \mathds R^{\nb}$ is given by
\begin{equation}\label{eq:obsv:FT:injection}
	\lb(\e_{\mathrm{b,y}}) = \begin{bNiceArray}{cccc:c:ccc} \nu_{1,1}\Phi_{1}^{l_1}(e_{\mathrm b,r_1}) & \cdots & \nu_{1,l_1-1}\Phi^{l_1-1}_{l_1}(e_{\mathrm b,r_1}) & \nu_{1,l_1}\Phi^{l_1}_{l_1}(e_{\mathrm b,r_1}) & \cdots & \nu_{\pb,1}\Phi_{1}^{l_{\pb}}(e_{\mathrm b,r_{\pb}}) & \cdots & \nu_{\pb,l_{\pb}}\Phi_{l_{\pb}}^{l_{\pb}}(e_{\mathrm b,r_{\pb}})
	\end{bNiceArray}^\transp,
\end{equation}
with positive parameters $\nu_{i,j}$, $i=1,\ldots,\pb$, $1\leq j \leq l_i$ and the nonlinear functions designed according to~\eqref{eq:diff:moreno:phi:1} and~\eqref{eq:diff:moreno:phi:2} in Section~\ref{sec:morenodiff}.
This allows to state the following 
\begin{theorem}\label{thm:FT:b}
For any constants $0<\mu<1$ and $-1<d_0 < 0 < d_\infty < \min_{i=1}^{\pb} \frac{1}{l_i-1}$, there exist appropriate gains $\nu_{i,j}>0$, $i=1,\ldots,\pb$, $j=1,\ldots,l_i$, such that~\eqref{eq:obsv:FT:b} with the output injection~\eqref{eq:obsv:FT:injection} is a continuous fixed-time observer for~\eqref{eq:sys_SO:b}, i.e. $\Sigma_b$.
\end{theorem}
\begin{remark}
By choosing $-1<d_0=d_\infty <0$ the error dynamics correspond to those of the homogeneous observer with finite time convergence proposed in~\cite{perruquetti2008finite}.
For $d_0=d_\infty=-1$, Levant's robust exact differentiator~\cite{levant1998robust} is obtained.
\end{remark}
\begin{proof}
	Following Section~\ref{sec:morenodiff}, the error dynamics of $\e_{\mathrm b} = \x_{\mathrm{b}}-\hat\x_{\mathrm{b}}$ are given by
	\begin{equation}
		\begin{aligned}
			\Sigma_{\mathrm e,\mathrm{b},1}:& \left\{\begin{aligned} %
				\dot e_{\mathrm b,1} &= e_{\mathrm b,2} - \nu_{1,1}\Phi_{1}^{l_1}(e_{\mathrm b,r_1})\\
				&\,\vdots \\
				\dot e_{\mathrm{b,l_1-1}} &= e_{\mathrm b, l_1} - \nu_{1,l_1-1}\Phi_{l_1-1}^{l_1}(e_{\mathrm b,r_1})\\
				\dot e_{\mathrm b,l_1} &= -\nu_{1,l_1}\Phi_{l_1}^{l_1}(e_{\mathrm b, r_1})
			\end{aligned}	\right. \\
			\Sigma_{\mathrm e,\mathrm{b},2}:& \left\{\begin{aligned} %
				\dot e_{\mathrm b,r_2} &= e_{\mathrm b,r_2+1} - \nu_{2,1}\Phi_{1}^{l_2}(e_{\mathrm b,r_2})\\
				&\,\vdots \\
				\dot e_{\mathrm{b,r_2+l_2-2}} &= e_{\mathrm b,r_2+l_2-1} - \nu_{2,l_2-1}\Phi_{l_2-1}^{l_2}(e_{\mathrm b,r_2})\\
				\dot e_{\mathrm b,r_2+l_2-1} &= -\nu_{2,l_2}\Phi_{l_2}^{l_2}(e_{\mathrm b,r_2})
			\end{aligned}	\right. \\
			&  \vdots \\
			\Sigma_{\mathrm e,\mathrm{b},\pb}:& \left\{\begin{aligned} 
				\dot e_{\mathrm b,r_{\pb}} &= e_{\mathrm b,r_{\pb}+1} - \nu_{\pb,1} \Phi_{1}^{l_{\pb}}(e_{\mathrm b,r_{\pb}})\\
				&\,\vdots \\
				\dot e_{\mathrm b,\nb-1} &= \dot e_{\mathrm b, \nb} - \nu_{\pb,l_{\pb}-1}\Phi_{l_{\pb}-1}^{l_{\pb}}(e_{\mathrm b, r_{\pb}}) \\
				\dot e_{\mathrm b,\nb} &= -\nu_{\pb,l_{\pb}}\Phi_{l_{\pb}}^{l_{\pb}}(e_{\mathrm b, r_{\pb}}) 
			\end{aligned}  \right.
		\end{aligned}
	\end{equation}
The systems $\Sigma_{\mathrm e, \mathrm b, i}$ for $i=1,\ldots,\pb$ are decoupled and each coincides with the error dynamics of Moreno's arbitrary order fixed-time estimator, i.e.~\eqref{eq:error:FT} with $u=0$. 
Hence, the result follows from~\cite[Theorem 1]{moreno2020arbitrary}.
\end{proof}
This result facilitates the following 
\begin{corollary}
	Suppose that the strongly observable system~\eqref{eq:sys_SO} is also strong$^*$ detectable and that $\hat \Sigma_{\mathrm b}$ in~\eqref{eq:obsv:FT:b} is designed according to Theorem~\eqref{thm:FT:b}.
	Then,~\eqref{eq:obsv:FT:b} together with~\eqref{eq:obsv:dred} is a continuous fixed-time unknown input observer for~\eqref{eq:sys_SO}.
\end{corollary}

If the system is strongly observable but not strong$^*$ detectable, derivatives of the output signals of subsystem $\Sigma_{\mathrm d}$ are required.
It is still possible, however, to achieve fixed-time convergence in this case by utilizing Moreno's fixed-time differentiator. 
Therefore, the following discontinuous bi-homogeneous observer (DBHO) is proposed for $\Sigma_{\mathrm d}$ in~\eqref{eq:sys_SO:d}:
	\begin{align}\label{eq:obsv:FT:d}
		\hat\Sigma_{\mathrm d}: \quad \dot{\hat \x}_{\mathrm{d}} &= (\Ad^\star + \Bd\Fdd) \hat\x_{\mathrm{d}} +\Bd\Fdb \hat{\x}_\mathrm{b} + \Bod\y_0 + \Hdd \yd + \ld(\e_{\mathrm{d,y}}),
	\end{align}	
where
\begin{equation}
	\e_{\mathrm{d,y}} = \yd - \hat\y_\mathrm{d} = \begin{bmatrix}
		e_{\mathrm d,w_1} & e_{\mathrm d,w_2} & \cdots & e_{\mathrm d,w_{\md}}
	\end{bmatrix}^\transp
\end{equation}
as the output error with $w_{i} = \sum_{j=1}^{i-1} q_j + 1$ and $i=1,2,\ldots,\md$.
Moreover, $\ld:\mathds R^{\md} \mapsto\mathds R^{\nd}$ is the nonlinear output injection with
\begin{equation}\label{eq:obsv:HO:injection}
	\ld(\e_{\mathrm{d,y}}) = \begin{bNiceArray}{cccc:c:ccc} \kappa_{1,1} \Phi_1^{q_1}(e_{\mathrm d,1}) & \cdots & \kappa_{1,q_1-1} \Phi_{q_1}^{q_1-1}(e_{\mathrm d,1})&\kappa_{1,q_1} \Phi_{q_1}^{q_1}(e_{\mathrm d,1}) & \cdots & \kappa_{\md,1} \Phi_{1}^{q_{\md}}(e_{\mathrm d,r_{\md}})  & \cdots &\kappa_{\md,q_{\md}} \Phi_{q_{\md}}^{q_{\md}}(e_{\mathrm d,r_{\md}}) 
	\end{bNiceArray}^\transp
\end{equation}
with positive parameters $\kappa_{i,j}$, $i=1,\ldots,\md$, $j=0,\ldots,q_{i}-1$ and the nonlinear functions $\Phi_j^{q_i}$ with $i=1,\ldots, \md$ and $j=1,\ldots,q_i$ ~\eqref{eq:diff:moreno:phi:1} and~\eqref{eq:diff:moreno:phi:2} in Section~\ref{sec:morenodiff}.

Based on this observer design, it is possible to achieve fixed-time convergence of the overall estimation error according to
\begin{theorem}\label{thm:FT:d}
	Suppose that $\hat \Sigma_{\mathrm b}$ in~\eqref{eq:obsv:FT:b} is designed according to Theorem~\ref{thm:FT:b} and that the unknown input $\ud$ is bounded.
	Then, for $d_0=-1$ and any constants $0 < d_\infty < \min_{i=1}^{\md} \frac{1}{q_i-1}$ and $0<\mu<1$, there exist appropriate gains $\kappa_{i,j}>0$, $i=1,\ldots,\md$, $1\leq j \leq q_i$, such that~\eqref{eq:obsv:FT:d} with the output injection~\eqref{eq:obsv:HO:injection} is a fixed-time observer for~\eqref{eq:sys_SO:d}, i.e. for $\Sigma_d$.
\end{theorem}
The proof follows from analogously to the proof of Theorem~\ref{THM:obs:sm}.
Hence, the combination of Theorems~\ref{thm:FT:b} and~\ref{thm:FT:d} allows to design a fixed-time unknown input observer for strongly observable systems by using ~\eqref{eq:obsv:FT:b} and~\eqref{eq:obsv:FT:d}.

\subsection{Summary of the Design Procedure}\label{sec:designsummary}

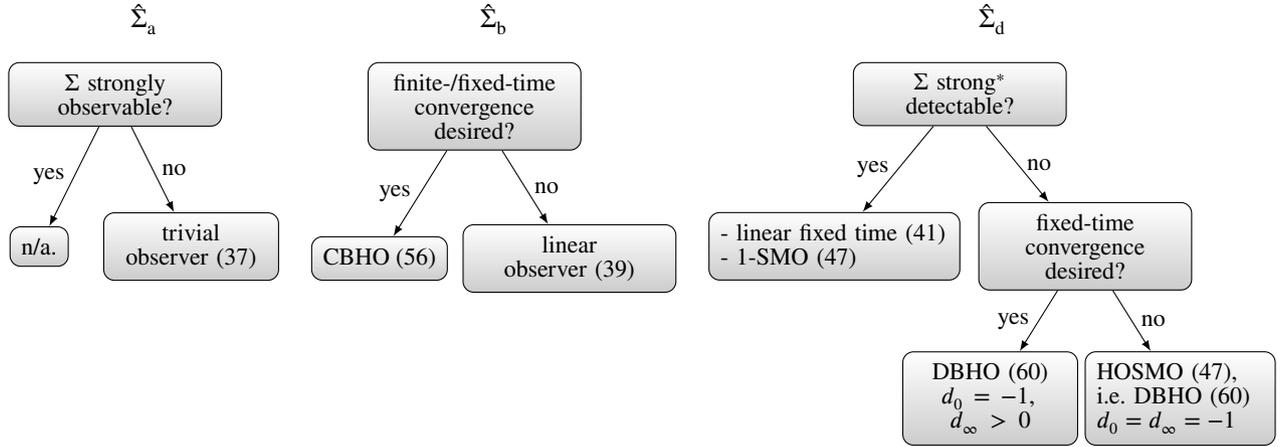
\begin{figure}\centering
	\begin{tabular}{ccc}
		$\hat \Sigma_\mathrm{a}$ & $\hat \Sigma_\mathrm{b}$ & $\hat \Sigma_\mathrm{d}$ \\
\begin{tikzpicture}[
	baseline=(current bounding box.north),
	every node/.style={font=\scriptsize},
	treenode/.style = {shape=rectangle, rounded corners,
		draw, align=center,
		top color=white, 
		bottom color=black!20, 
		inner sep=1.5mm, 
		},
	grow=south, level distance=2cm,
	level 1/.style={sibling distance=2.0cm},
	level 2/.style={sibling distance=5cm},
	edge from parent/.style = {draw, -latex},
	]
	
	\node [treenode, text width = 2.5cm] {$\Sigma$ strongly observable?}
	child {node [treenode] {n/a.} edge from parent node [left] {yes}}
	child { node [treenode,text width = 2cm] {trivial observer~\eqref{eq:obsv:a}}
		edge from parent node [right] {no} }
	;
	
\end{tikzpicture} & 
\begin{tikzpicture}[
	baseline=(current bounding box.north),
every node/.style={font=\scriptsize},
treenode/.style = {shape=rectangle, rounded corners,
	draw, align=center,
	top color=white, 
	bottom color=black!20, 
	inner sep=1.5mm, 
},
grow=south, level distance=2cm,
level 1/.style={sibling distance=2.5cm},
level 2/.style={sibling distance=5cm},
edge from parent/.style = {draw, -latex},
]

\node [treenode, text width = 2.5cm] {finite-/fixed-time convergence desired?}
child { node [treenode] {CBHO~\eqref{eq:obsv:FT:b}}
	edge from parent node [left] {yes} }
child {node [treenode, text width = 2.5cm] {linear observer \eqref{eq:obsv:b}} edge from parent node [right] {no}};

\end{tikzpicture} &
\begin{tikzpicture}[
	baseline=(current bounding box.north),
	every node/.style={font=\scriptsize},
	treenode/.style = {shape=rectangle, rounded corners,
		draw, align=center,
		top color=white, 
		bottom color=black!20, 
		inner sep=1.5mm, 
	},
	grow=south, level distance=2.0cm,
	level 1/.style={sibling distance=3.3cm},
	level 2/.style={sibling distance=2.5cm},
	edge from parent/.style = {draw, -latex},
	]
	
	\node [treenode, text width = 2.5cm] {$\Sigma$ strong$^*$ detectable?}
	child { node [treenode, text width = 3cm,align=left] {
				- linear fixed time \eqref{eq:obsv:dred}\\
				- 1-SMO~\eqref{eq:obs:d:SM}}
			  		edge from parent node [left] {yes} }
	child {node [treenode, text width = 2.5cm] {fixed-time convergence desired?} 
		child {node[treenode,text width = 2cm] {DBHO~\eqref{eq:obsv:FT:d}\\ $d_0=-1,$\, \\$d_\infty>0$}
			edge from parent node [left] {yes}} 
		child[-] {node[treenode,align=left,,text width = 2.2cm] {HOSMO \eqref{eq:obs:d:SM}, \\ i.e. DBHO~\eqref{eq:obsv:FT:d}\\ $d_0=d_\infty=-1$ 
			                      }
	edge from parent node [right] {no}} 
    edge from parent node [right] {no}
};

\end{tikzpicture} 
	\end{tabular}
\caption{Observer design for the specific subsystems depending on the observability properties and the convergence requirements.}
\label{fig:designtree}
\end{figure}

This section summarizes the design procedure and discusses important design aspects.
In general, it is possible to combine various observer design approaches from the previous sections, which yields a variety of observers. 
It should be remarked that the discontinuous and continuous bi-homogeneous observers presented in Section~\ref{sec:finite/fixed} can achieve either finite-time or fixed-time convergence by a proper selection of the homogeneity degrees $d_0$ and $d_\infty$~\cite{moreno2020arbitrary}. 
Hence, both convergence properties for these observers are considered.
The choice for one specific observer is a design question.
The proposed design techniques are summarized in Table~\ref{tab:design_summary} and Fig.~\ref{fig:designtree}.

If the system is not strongly observable, $\Sigma_\mathrm{a}$ exists and a trivial observer $\hat\Sigma_\mathrm{a}$ according to~\eqref{eq:obsv:a} has to be designed.

For subsystem (b), either asymptotic convergence utilizing a linear observer according to~\eqref{eq:obsv:b} or finite-/fixed time convergence can be achieved. 
For the latter, a bi-homogeneous observer with continuous right-hand side (CBHO) according to~\eqref{eq:obsv:FT:b} has to be designed.
There, the homogeneity degree $d_0$ has to be chosen according to $-1<d_0<0$.

The observer design for subsystem (d) mainly depends on the strong$^*$ detectability property.
If the system is strong$^*$ detetable, it is possible to choose the linear fixed-time observer~\eqref{eq:obsv:dred} or, if one is interest in an estimate of the unknown input, it is also possible to design a first order SMO (1-SMO) according to~\eqref{eq:obs:d:SM} and Remark~\ref{remark:1smo}.
If subsystem (d) is solely strongly detectable, derivatives of the outputs of subsystem (d) are required.
The number of performed differentiations are kept at a minimum.
The proposed design allows to achieve fixed time convergence with a discontiunous bi-homogeneous observer (DBHO) according to~\eqref{eq:obsv:FT:d} with the homogeneity degrees $d_0=-1$ and $0<d_\infty< \frac{1}{q_i-1}$ for every integrator chain in subsystem (d). 
If finite time convergence is desired, the same observer with homogeneity degrees $d_0 =-1$ and $-1 \leq d_\infty< 0$ can be employed.
For  $d_0=d_\infty=-1$ this results in the sliding mode observer~\eqref{eq:obs:d:SM}.

\begin{table}
	\caption{Summary of observer designs for each subsystem together with the convergence properties. (CBHO $\ldots$ continuous bi-homogeneous observer, DBHO $\ldots$ discontinuous bi-homogeneous observer, 1-SMO $\ldots$ first order sliding mode observer, HOSMO $\ldots$ higher order sliding mode observer)}
	\label{tab:design_summary}
	\centering
	\begin{tabular}{ccccc}
		\toprule
		& strong$^*$ det. & strongly det. & strongly obsv. & strongly obsv. + strong$^*$ det.\\
		\midrule
		$\hat\Sigma_\mathrm{a}$	&  linear (asymptotic) & linear (asymptotic) & n/a. & n/a.\\ \midrule 
		$\hat\Sigma_\mathrm{b}$ &\begin{tabular}{@{}c@{}}linear (asymptotic) \\ CBHO (finite-/fixed-time)\end{tabular} & \begin{tabular}{@{}c@{}}linear (asymptotic) \\CBHO (finite-/fixed-time)\end{tabular} & \begin{tabular}{@{}c@{}}linear (asymptotic) \\ CBHO (finite-/fixed-time)\end{tabular} & \begin{tabular}{@{}c@{}}linear (asymptotic) \\ CBHO (finite-/fixed-time)\end{tabular}	\\ \midrule
		$\hat\Sigma_\mathrm{d}$ & \begin{tabular}{@{}c@{}}linear (fixed-time)\\ 1-SMO (finite-time)\end{tabular} &  \begin{tabular}{@{}c@{}}HOSMO (finite-time) \\ DBHO (finite-/fixed-time)\end{tabular} & \begin{tabular}{@{}c@{}}HOSMO (finite-time) \\ DBHO (finite/fixed-time) \end{tabular} & \begin{tabular}{@{}c@{}}linear (fixed-time)\\ 1-SMO (finite-time)\end{tabular}\\
		\bottomrule
	\end{tabular} 
\end{table}

\section{Example}\label{sec:example}

As a numerical example, system~\eqref{eq:sys:orig} is considered with the following coefficient matrices:
\begin{align}\label{eq:example}
	\A &= \left[\begin{array}{*{7}{r@{\hspace{3ex}}}r}
		 -4  &   6  &  -5  &  11  &   8  &  -5  &  22  &   0 \\
		 -3  &   0  &   2  &  -2  & -1  &   1  &  -3  & -12 \\
		 -3  &  -2  &   3  &  -9  &  -2 &    6 &   -8 &   -2 \\
  		0  &   0   &  0  &  -2  &   1  &   0 &   -2  &   0 \\
		3  &   0   & -1&     3 &    1  &   0  &   4 &   11\\
		3  &   1 &   -4 &    6 &    1 &   -7  &   3  &   2\\
		0  &   0 &    0 &   -1  &   0 &    0 &   -1   &  1\\
		-3 &    0&     1 &   -5&    -1 &    1&    -7  & -10
	\end{array} \right],\qquad
	\B = \left[\begin{array}{r@{\hspace{3ex}}r}
	0   & -2 \\
0     &1 \\
-1    & 1 \\
0     &0 \\
0    &-1 \\
1    &-1 \\
0    & 0 \\
0    & 1
\end{array} \right], \\ 
\C &= \left[\begin{array}{*{7}{r@{\hspace{3ex}}}r}
0   &  0  &   0  &   1  &   0  &   0    & 1  &   0 \\
0   &  0  &   0  &   0   &  0&     0    & 1 &    0 \\
0   & -2 &    1  &   0 &   -2   &  1   & -2 &    0
\end{array} \right], \qquad \D = \bm 0.
\end{align}
It can be verified that $\rank \C\B = 0$ and hence a linear UIO does not exist.
A transformation to the SCB~\eqref{eq:scb} using the linear systems toolkit\footnote{\Matlab code available via \href{http://www.mae.cuhk.edu.hk/~bmchen/}{http://www.mae.cuhk.edu.hk/~bmchen/}} and the transformation\footnote{\Matlab code available via \href{http://www.reichhartinger.at/index.php?id=38}{http://www.reichhartinger.at/index.php?id=38}} in Theorem~\ref{THM:TRANSF_SUBSYS_D} reveals that $\na=1$, $\nb=2$, $\nc=0$ and $\nd =5$.
For subsystem (d), there are two chains of integrators with $q_1=3$ and $q_2 =2$, respectively.
This (unstable) system is strongly detectable, because it has a stable invariant zero at $\lambda_1 = -10$ and $n_c= 0$. The transformation to the proposed SCB~\eqref{eq:scb} with $\Fdd$ as in~\eqref{eq:Fdd} results in the transformation matrix
\begin{equation}
	\Ts =    \begin{bmatrix}
   -1 &    -0.3297  & 0.0330   & 5.6804 &  -1.0328   &      0 &   3.7947 &  -1.2649\\
0  & -0.2857 &  -0.1429&    0.7746 &  -0.7746   &      0   &-2.2136 &   0.6325 \\
0 &   0.4286&   -0.2857 & -11.8771 &   4.6476&   -1.2910 &  -0.3162 &   0.9487\\
0 &        0 &        0 &   0.7746 &        0&         0&   -0.6325 &        0 \\
0 &       0  &       0 &  -1.2910  &  0.7746 &        0 &   1.5811 &  -0.6325\\
0 &        0 &        0&   11.8771 &  -4.6476 &   1.2910 &   0.3162&   -0.9487\\
0 &        0&         0 &   0.5164 &        0 &        0 &   0.6325&         0 \\
0 &        0 &        0 &  -2.8402 &   0.5164 &        0&   -1.8974 &   0.6325
	\end{bmatrix}.
\end{equation}
The transformed system in the SCB is given by
\begin{subequations}
\begin{align}
	\Aa &= -10,\quad \Ab = \begin{bmatrix}
		-1&   1\\
		-1  & 0\\ 
	\end{bmatrix},\quad \Cb = [ 1  \;0],\\
\Hab &= 0,\quad \Had = [ -0.0709\;  -0.3823], \quad \Hbd =  
\begin{bmatrix}       -4.3894  & -0.6325 \\
	8.7788   & 1.2649 	
\end{bmatrix}\\
\Fda &= \begin{bmatrix}
	    1.1619\\
	4.7434
\end{bmatrix},\quad \Fdb = \begin{bmatrix} 
-0.6682   & 0.4043 \\
2.2414   &-0.6081\end{bmatrix},\quad 
\Hdd = \begin{bmatrix}
 -6.2000   &   -0.2449   \\
-15.6000   &      -3.1843  \\
-21.5000   &     -4.8990    \\
-1.4697    &     -2.8000   \\
-11.3493   &     0.6000    \\
\end{bmatrix},\quad \Fdd = \begin{bmatrix}
0 & 0& 0& 0& 0 \\
0 & 0& -0.8165 & 0 & 0 
\end{bmatrix},\\
\Ad^\star &=\begin{bmatrix} 
  0 &   1  &      0  &       0 &        0\\
	0      &   0   & 1 &         0     &    0\\
	0     &    0  &       0    &     0    &     0 \\
	0   &      0 &        0 &        0 &   1\\
	0    &     0 &  0 &        0    &     0 \end{bmatrix},
\quad
\Bd = \begin{bmatrix}
	 0   &  0 \\
	0    & 0\\
	1    & 0\\
	0    & 0\\
	0    & 1
\end{bmatrix},\quad \Cd = \begin{bmatrix}
     1  &   0 &    0 &    0 &    0 \\
0   &  0 &    0  &   1 &    0
\end{bmatrix}.
\end{align}	
\end{subequations}
with the input and output transformation matrices are given by
\begin{equation}\label{eq:iotransf}
\Ti = \begin{bmatrix}
	 1.2910 &  -0.3162 \\
	0   & 0.6325 
\end{bmatrix} \quad \text{and} \quad \To^{-1} = \begin{bmatrix}
    0.7746  &  0 &   0 \\
-0.6325  &  1.5811&   0 \\
0 &    0 &   1 
\end{bmatrix}.
\end{equation}
The two components of the unknown input are chosen as $u_1(t)=1.5\sin(t)+0.5$ and $u_2(t)=\sigma(t-1) - \sigma(t-4)$, where $\sigma(t)$ is the unit step function. 
The components of the unknown input are thus bounded according to $u_1 \in [-2,\; 1]$ and $u_2 \in [0,\; 1]$.
Together with~\eqref{eq:iotransf} and~\eqref{eq:UIbounds}, this results in the input bounds $|u_{d,1}| \leq 1.5492=\Delta_{\mathrm d,1}$ and $|u_{d,2}| \leq 1.5811=\Delta_{\mathrm d,2}$ in the SCB.

Following Section~\ref{sec:observerdesign}, a sliding mode based UIO is designed for the strongly detectable system.
For subsystem (a), a trivial observer 
\begin{equation}\label{eq:obsvexample:a}
	\hat\Sigma_{\mathrm a}:\;	\dot{\hat\x}_\mathrm{a} = \Aa {\hat\x}_\mathrm{a} + \Hab \yb+\Had \yd 
\end{equation}
is employed.
For subsystem (b), a Luenberger observer
\begin{equation}\label{eq:obsvexample:b}
	\hat\Sigma_{\mathrm b}:\;	\dot{\hat\x}_\mathrm{b} = \Ab \hat\x_\mathrm{b} + \Hbd\yd +\Lb(\yb-\Cb\hat\x_\mathrm{b})
\end{equation}
is designed such that the eigenvalues of $(\Ab-\Lb\Cb)$ are given by the set $\{-8,-6\}$, which results in $\Lb=[13\;\, 47]^\transp$.
For subsystem (d), the sliding mode observer proposed in Section~\ref{sec:asymptotic_SMO} is chosen according to
\begin{subequations}\label{eq:obsexample:d:SM}
	\begin{align}
		\hat\Sigma_{\mathrm d}: \quad \dot{\hat \x}_{\mathrm{d}} &= (\Ad^\star + \Bd\Fdd) \hat\x_{\mathrm{d}} +\Bd(\Fda \hat\x_\mathrm{a}+\Fdb \hat{\x}_\mathrm{b})+ \Hdd \yd + \ld(\e_{\mathrm{d,y}}),
	\end{align}	
\end{subequations}
where
\begin{equation}
	\e_{\mathrm{d,y}} = \yd - \hat\y_\mathrm{d} = \begin{bmatrix}
		e_{\mathrm d,1} & e_{\mathrm d,4}
	\end{bmatrix}^\transp
\end{equation}
is the output error.
Moreover, $\ld:\mathds R^{2} \mapsto\mathds R^{5}$ is the nonlinear output injection with
\begin{equation}
	\ld(\e_{\mathrm{d,y}}) = \begin{bmatrix} \kappa_{1,1} \lfloor e_{\mathrm d,1}\rceil^{\frac{2}{3}} & \kappa_{1,2} \lfloor e_{\mathrm d,1} \rceil^{\frac{1}{3}}&\kappa_{1,3} \lfloor e_{\mathrm d,1}\rceil^0 &  \kappa_{2,1} \lfloor e_{\mathrm d,4} \rceil^{\frac{1}{2}}  & \kappa_{2,2} \lfloor e_{\mathrm d,4}\rceil^0
	\end{bmatrix}^\transp
\end{equation}
and 
\begin{equation}
	\Phi_j^{q_i}(z) = \lfloor z \rceil^{\frac{q_i-j}{q_i}}\;\text{for}\; i=1,\ldots,\md;\,j=1,\ldots,q_i.
\end{equation}
The gains for the sliding mode observer are chosen according to 
\begin{align*}
	\kappa_{1,1} &=2\Delta_{\mathrm d,1}^{\frac{1}{3}} , \quad &\kappa_{1,2} &= 2.12\Delta_{\mathrm d,1}^{\frac{2}{3}},\quad &\kappa_{1,3} &= 1.1\Delta_{\mathrm d,1}, \\
	\kappa_{2,1} &= 1.5\Delta_{\mathrm d,2}^{\frac{1}{2}}, \quad& \kappa_{2,2} &= 1.1\Delta_{\mathrm d,2},
\end{align*}
see also~\cite[Chapter 6]{shtessel2013sliding}.

\begin{figure}[tbp]
	\centering
	\includegraphics[width=0.95\linewidth]{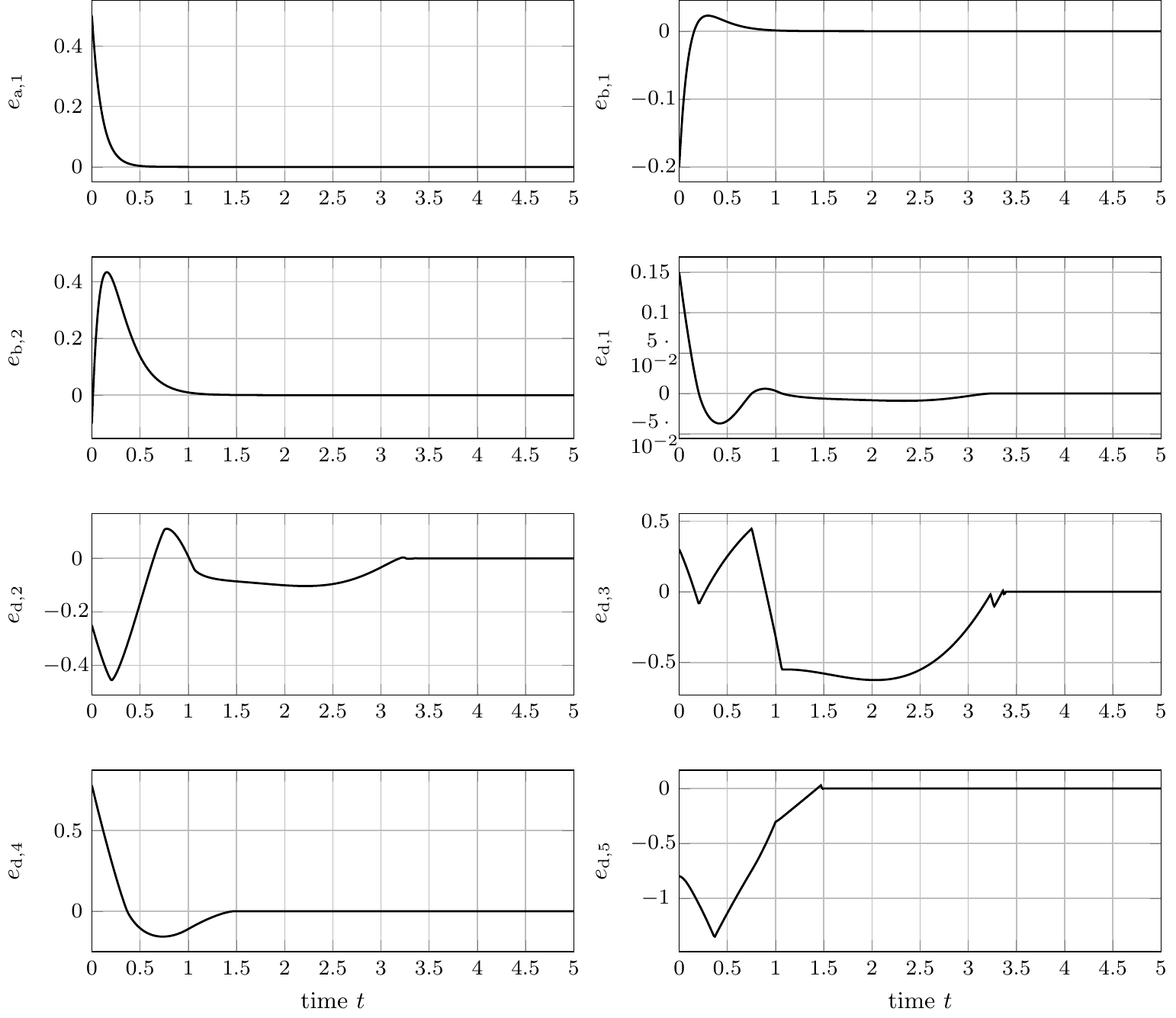}
	\caption{Estimation errors in SCB coordinates.}
	\label{fig:ebar}
\end{figure}

\begin{figure}[tbp]
	\centering
	\includegraphics[width=0.95\linewidth]{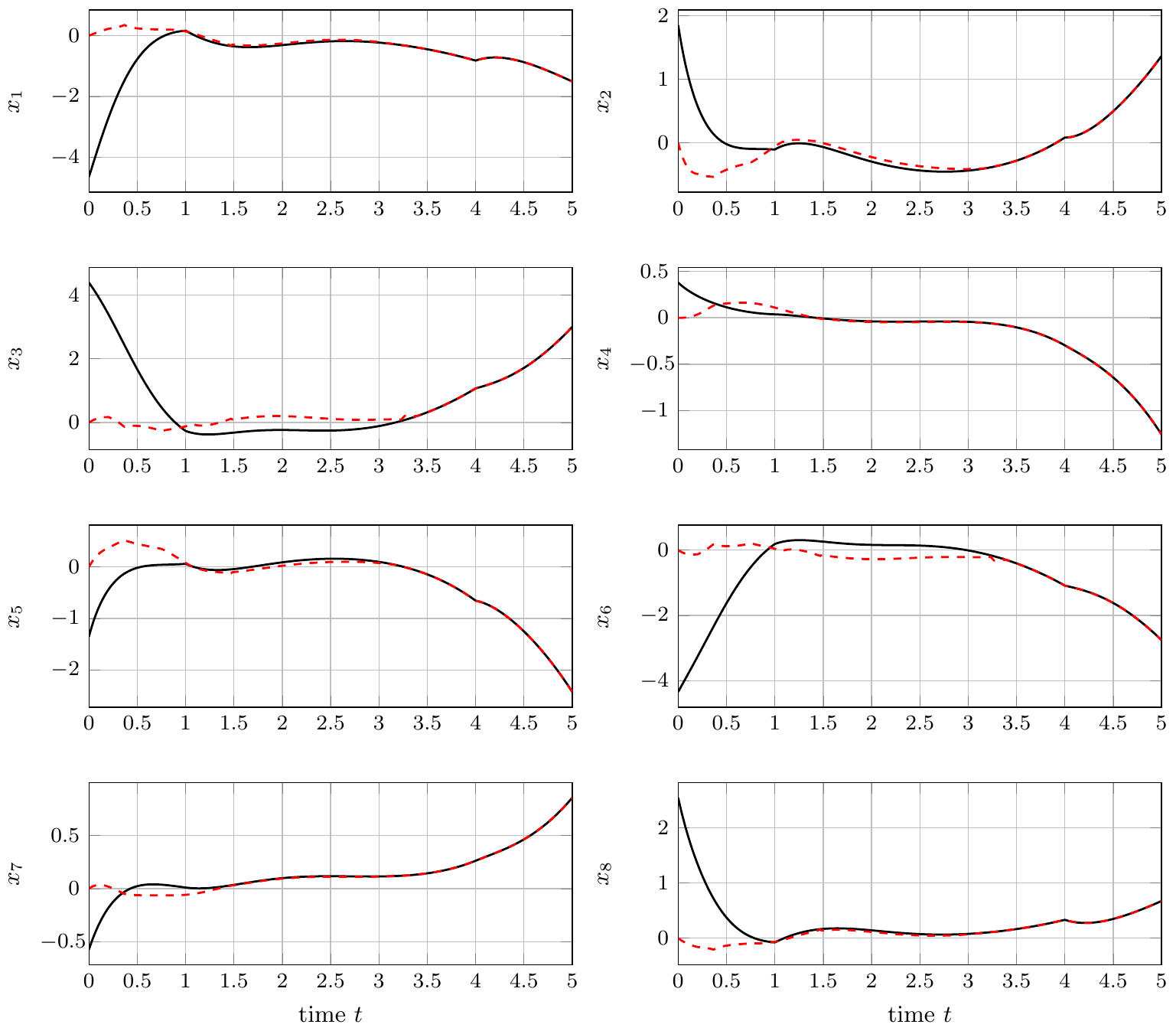}
	\caption{True (solid) and estimated system states (dashed red) for the original system~\eqref{eq:example}.}
	\label{fig:states}
\end{figure}

The initial condition of system~\eqref{eq:sys:orig} is chosen as $\x_0 = \begin{bmatrix}1& -0.1& 2& -0.3& 0.5& 0.2& -0.5& -0.2 \end{bmatrix}^\transp$ and the observer is initialized with zero.
The components of the estimation error in SCB are depicted in Fig.~\ref{fig:ebar}.
Here, the finite-time convergence properties of the errors in subsystem (d) can be verified.
Fig.~\ref{fig:states} shows a comparison of the states and their corresponding estimates in the original coordinates.
It can be seen that the estimates (asymptotically) converge to the true states.

\section{Discussion and Outlook}\label{sec:discussion}

This paper presents a unifying design framework for linear and nonlinear unknown input observers for linear time-invariant systems.
It is shown that after the transformation to the special coordinate basis, the design for asymptotic and finite- or fixed-time can be carried out in a similar fashion.
Depending on the system properties and the desired estimation error dynamics, the design procedure allows a straightforward design of asymptotic, finite- or fixed-time unknown input observers.
If derivatives are required, the number of differentiation operations is kept at a minimum. 
The design can be straightforwardly extended to descriptor systems~\cite{chen2004linear,bejarano2013observability} or unbounded unknown inputs \cite{bejarano2010high-order}.

In future work, the proposed observer design will be extensively evaluated in simulation studies and real-world experiments.
The proposed method could also be employed to reconstruct the unknown input by following the ideas in~\cite{niederwieser2021high} and it can hence be utilized in a robust control framework.
For this, the performance in a control loop needs thorough investigation.
Moreover, the influence of measurement noise and model uncertainty will be investigated.

\bibliographystyle{WileyNJD-AMA}
\bibliography{wileyNJD-AMA}%

\begin{thebibliography}{10}
\providecommand \doibase [0]{http://dx.doi.org/}%

\bibitem{bhattacharyya1978observer}
Bhattacharyya S. Observer design for linear systems with unknown inputs. {\it
  {IEEE} Transactions on Automatic Control} 1978\string; 23(3)\string:
  483--484.
\newblock \href {\doibase 10.1109/tac.1978.1101758} {doi:
  10.1109/tac.1978.1101758}

\bibitem{molinari1976astrong}
Molinari B. A strong controllability and observability in linear multivariable
  control. {\it {IEEE} Transactions on Automatic Control} 1976\string;
  21(5)\string: 761--764.
\newblock \href {\doibase 10.1109/tac.1976.1101364} {doi:
  10.1109/tac.1976.1101364}

\bibitem{hautus1983strong}
Hautus M. Strong detectability and observers. {\it Linear Algebra and its
  Applications} 1983\string; 50\string: 353--368.
\newblock \href {\doibase 10.1016/0024-3795(83)90061-7} {doi:
  10.1016/0024-3795(83)90061-7}

\bibitem{hou1994fault}
Hou M, Müller PC. Fault detection and isolation observers. {\it International
  Journal of Control} 1994\string; 60(5)\string: 827--846.
\newblock \href {\doibase 10.1080/00207179408921497} {doi:
  10.1080/00207179408921497}

\bibitem{kratz1995characterization}
Kratz W. Characterization of strong observability and construction of an
  observer. {\it Linear Algebra and its Applications} 1995\string; 221\string:
  31--40.
\newblock \href {\doibase 10.1016/0024-3795(93)00221-k} {doi:
  10.1016/0024-3795(93)00221-k}

\bibitem{valcher1999state}
Valcher M. State observers for discrete-time linear systems with unknown
  inputs. {\it {IEEE} Transactions on Automatic Control} 1999\string;
  44(2)\string: 397--401.
\newblock \href {\doibase 10.1109/9.746275} {doi: 10.1109/9.746275}

\bibitem{chen1999robust}
Chen J, Patton R. {\it Robust Model-Based Fault Diagnosis for Dynamic Systems}.
\newblock Boston, MA: Springer US .
\newblock 1999.

\bibitem{alwi2008fault}
Alwi H, Edwards C. Fault Detection and Fault-Tolerant Control of a Civil
  Aircraft Using a Sliding-Mode-Based Scheme. {\it {IEEE} Transactions on
  Control Systems Technology} 2008\string; 16(3)\string: 499--510.
\newblock \href {\doibase 10.1109/tcst.2007.906311} {doi:
  10.1109/tcst.2007.906311}

\bibitem{ferreira2011robust}
Ferreira A, Bejarano FJ, Fridman LM. Robust Control With Exact Uncertainties
  Compensation: With or Without Chattering?. {\it {IEEE} Transactions on
  Control Systems Technology} 2011\string; 19(5)\string: 969--975.
\newblock \href {\doibase 10.1109/tcst.2010.2064168} {doi:
  10.1109/tcst.2010.2064168}

\bibitem{saif1992decentralized}
Saif M, Guan Y. Decentralized state estimation in large-scale interconnected
  dynamical systems. {\it Automatica} 1992\string; 28(1)\string: 215--219.
\newblock \href {\doibase 10.1016/0005-1098(92)90024-a} {doi:
  10.1016/0005-1098(92)90024-a}

\bibitem{taha2015unknown}
Taha AF, Elmahdi A, Panchal JH, Sun D. Unknown input observer design and
  analysis for networked control systems. {\it International Journal of
  Control} 2015\string: 1--15.
\newblock \href {\doibase 10.1080/00207179.2014.985718} {doi:
  10.1080/00207179.2014.985718}

\bibitem{edwards1998sliding}
Edwards C. {\it Sliding mode control : {T}heory and applications}.
\newblock London: Taylor \& Francis .
\newblock 1998.

\bibitem{edwards2006acomparison}
Edwards C, Tan CP. A Comparison of Sliding Mode and Unknown Input Observers for
  Fault Reconstruction. {\it European Journal of Control} 2006\string;
  12(3)\string: 245--260.
\newblock \href {\doibase 10.3166/ejc.12.245-260} {doi: 10.3166/ejc.12.245-260}

\bibitem{alwi2011fault}
Alwi H, Edwards C, Tan CP. {\it Fault Detection and Fault-Tolerant Control
  Using Sliding Modes}.
\newblock Springer-Verlag GmbH .
\newblock 2011.

\bibitem{tranninger2019exact}
Tranninger M, Seeber R, Steinberger M, Horn M. Exact State Reconstruction for
  {LTI}-Systems with Non-Differentiable Unknown Inputs. In: 18th European
  Control Conference ({ECC}). {IEEE}; 2019\string: 3096--3102

\bibitem{levant1998robust}
Levant A. Robust exact differentiation via sliding mode technique. {\it
  Automatica} 1998\string; 34(3)\string: 379--384.
\newblock \href {\doibase 10.1016/s0005-1098(97)00209-4} {doi:
  10.1016/s0005-1098(97)00209-4}

\bibitem{moreno2020arbitrary}
Moreno JA. Arbitrary Order Fixed-Time Differentiators. {\it {IEEE} Transactions
  on Automatic Control} 2021\string; accepted for publication\string: 1--1.
\newblock \href {\doibase 10.1109/tac.2021.3071027} {doi:
  10.1109/tac.2021.3071027}

\bibitem{bejarano2007exact}
Bejarano F, Fridman L, Poznyak A. Exact state estimation for linear systems
  with unknown inputs based on hierarchical super-twisting algorithm. {\it
  International Journal of Robust and Nonlinear Control} 2007\string;
  17(18)\string: 1734--1753.
\newblock \href {\doibase 10.1002/rnc.1190} {doi: 10.1002/rnc.1190}

\bibitem{bejarano2009unknown}
Bejarano FJ, Fridman L, Poznyak A. Unknown Input and State Estimation for
  Unobservable Systems. {\it {SIAM} Journal on Control and Optimization}
  2009\string; 48(2)\string: 1155--1178.
\newblock \href {\doibase 10.1137/070700322} {doi: 10.1137/070700322}

\bibitem{fridman2011high-order}
Fridman L, Davila J, Levant A. High-order sliding-mode observation for linear
  systems with unknown inputs. {\it Nonlinear Analysis: Hybrid Systems}
  2011\string; 5(2)\string: 189--205.
\newblock \href {\doibase 10.1016/j.nahs.2010.09.003} {doi:
  10.1016/j.nahs.2010.09.003}

\bibitem{bejarano2010high-order}
Bejarano FJ, Fridman L. High order sliding mode observer for linear systems
  with unbounded unknown inputs. {\it International Journal of Control}
  2010\string; 83(9)\string: 1920--1929.
\newblock \href {\doibase 10.1080/00207179.2010.501386} {doi:
  10.1080/00207179.2010.501386}

\bibitem{niederwieser2019ageneralization}
Niederwieser H, Koch S, Reichhartinger M. A Generalization of Ackermann's
  Formula for the Design of Continuous and Discontinuous Observers. In: IEEE.
  {IEEE}; 2019

\bibitem{niederwieser2021high}
Niederwieser H, Tranninger M, Seeber R, Reichhartinger M. Higher-order sliding
  mode observer design for linear time-invariant multivariable systems based on
  a new observer normal form. {\it arXiv preprint} 2021.

\bibitem{sannuti1987special}
Sannuti P, Saberi A. Special coordinate basis for multivariable linear
  systems{\textemdash}finite and infinite zero structure, squaring down and
  decoupling. {\it International Journal of Control} 1987\string; 45(5)\string:
  1655--1704.
\newblock \href {\doibase 10.1080/00207178708933840} {doi:
  10.1080/00207178708933840}

\bibitem{chen2004linear}
Chen BM, Lin Z, Shamash Y. {\it {L}inear {S}ystems {T}heory}.
\newblock Birkhäuser Boston .
\newblock 2004

\bibitem{saberi1990squaring}
Sannuti aAS. Squaring down of non-strictly proper systems. {\it International
  Journal of Control} 1990\string; 51(3)\string: 621--629.
\newblock \href {\doibase 10.1080/00207179008934088} {doi:
  10.1080/00207179008934088}

\bibitem{ozcetin1990special}
Ozcetin HK, Saberi A, Sannuti P. Special coordinate basis for order reduction
  of linear multivariable systems. {\it International Journal of Control}
  1990\string; 52(1)\string: 191--226.
\newblock \href {\doibase 10.1080/00207179008953531} {doi:
  10.1080/00207179008953531}

\bibitem{chen2011loop}
Chen BM, Saberi A, Sannuti P. {\it Loop Transfer Recovery: Analysis and
  Design}.
\newblock Springer London .
\newblock 2011.

\bibitem{chen1993construction}
Chen BM, Saberi A, Sannuti P, Shamash Y. Construction and parameterization of
  all static and dynamic {{$H_2$}}-optimal state feedback solutions, optimal
  fixed modes and fixed decoupling zeros. {\it {IEEE} Transactions on Automatic
  Control} 1993\string; 38(2)\string: 248--261.
\newblock \href {\doibase 10.1109/9.250513} {doi: 10.1109/9.250513}

\bibitem{chen2000robust}
Chen BM. {\it Robust and {{$H_\infty$}} control}.
\newblock London New York: Springer .
\newblock 2000.

\bibitem{chu2002onthenumerical}
Chu D, Liu X, Tan R. On the numerical computation of a structural decomposition
  in systems and control. {\it {IEEE} Transactions on Automatic Control}
  2002\string; 47(11)\string: 1786--1799.
\newblock \href {\doibase 10.1109/tac.2002.804484} {doi:
  10.1109/tac.2002.804484}

\bibitem{xiong1999functional}
Xiong Y, Saif M. Functional observers for linear systems with unknown inputs.
  {\it {IFAC} Proceedings Volumes} 1999\string; 32(2)\string: 1832--1837.
\newblock \href {\doibase 10.1016/s1474-6670(17)56311-9} {doi:
  10.1016/s1474-6670(17)56311-9}

\bibitem{xiong1999robust}
Xiong Y, Saif M. Robust fault isolation observer design. {\it Proceedings of
  the 1999 American Control Conference (Cat. No. 99CH36251)} 1999.
\newblock \href {\doibase 10.1109/acc.1999.786285} {doi:
  10.1109/acc.1999.786285}

\bibitem{saif2003sliding}
Saif M, Xiong Y. {\it Sliding Mode Observers and Their Application in Fault
  Diagnosis}ch.~1\string: 1--57; Springer Berlin Heidelberg .
\newblock 2003

\bibitem{filippov1988differential}
Filippov AF. {\it Differential Equations with Discontinuous Righthand Sides}.
\newblock Springer Netherlands .
\newblock 1988.

\bibitem{trentelman2012control}
Trentelman H, Stoorvogel AA, Hautus ML. {\it Control Theory for Linear
  Systems}.
\newblock Springer London .
\newblock 2012.

\bibitem{skogestad2005multivariable}
Skogestad P. {\it Multivariable Feedback Control: Analysis and Design}.
\newblock John Wiley \& Sons .
\newblock 2005.

\bibitem{andrieu2008homogeneous}
Andrieu V, Praly L, Astolfi A. Homogeneous Approximation, Recursive Observer
  Design, and Output Feedback. {\it {SIAM} Journal on Control and Optimization}
  2008\string; 47(4)\string: 1814--1850.
\newblock \href {\doibase 10.1137/060675861} {doi: 10.1137/060675861}

\bibitem{levant2003high-order}
Levant A. Higher-order sliding modes, differentiation and output-feedback
  control. {\it International Journal of Control} 2003\string; 76(9-10)\string:
  924--941.
\newblock \href {\doibase 10.1080/0020717031000099029} {doi:
  10.1080/0020717031000099029}

\bibitem{shtessel2013sliding}
Shtessel Y, Edwards C, Fridman L, Levant A. {\it Sliding Mode Control and
  Observation}.
\newblock Springer-Verlag GmbH .
\newblock 2013.

\bibitem{liu2005linear}
Liu X, Chen BM, Lin Z. Linear systems toolkit in Matlab: structural
  decompositions and their applications. {\it Journal of Control Theory and
  Applications} 2005\string; 3(3)\string: 287--294.
\newblock \href {\doibase 10.1007/s11768-005-0051-0} {doi:
  10.1007/s11768-005-0051-0}

\bibitem{morse1973structural}
Morse AS. Structural Invariants of Linear Multivariable Systems. {\it {SIAM}
  Journal on Control} 1973\string; 11(3)\string: 446--465.
\newblock \href {\doibase 10.1137/0311037} {doi: 10.1137/0311037}

\bibitem{perruquetti2008finite}
Perruquetti W, Floquet T, Moulay E. Finite-Time Observers: Application to
  Secure Communication. {\it {IEEE} Transactions on Automatic Control}
  2008\string; 53(1)\string: 356--360.
\newblock \href {\doibase 10.1109/tac.2007.914264} {doi:
  10.1109/tac.2007.914264}

\bibitem{bejarano2013observability}
Bejarano FJ, Floquet T, Perruquetti W, Zheng G. Observability and detectability
  of singular linear systems with unknown inputs. {\it Automatica} 2013\string;
  49(3)\string: 793--800.
\newblock \href {\doibase 10.1016/j.automatica.2012.11.043} {doi:
  10.1016/j.automatica.2012.11.043}

\end{thebibliography}

\appendix
\section{Proof of Lemma {{\ref{LE:SCB_SPECIAL}}}}\label{sec:proof_of_theorem}
Due to the special structure of $\T$ it is sufficient to proof the existence of a transformation $\x_\mathrm{d} = \Td^{-1} \bar \x_\mathrm{d}$ such that subsystem (d) takes the desired form as in Proposition~\ref{prop:SCB} with $\Fdd$ as in~\eqref{eq:Fdd}. 
The existence of such a transformation for strongly observable systems follows from~\cite[Theorem 3.1]{niederwieser2021high}.
It is noted that subsystem (d), i.e. the triple $(\Ad,\Bd,\Cd$), is strongly observable, because it possesses no invariant zeros.
The constructive proof of~\cite[Theorem 3.1]{niederwieser2021high}, i.e., the transformation algorithm presented in \cite[Section 4.1]{niederwieser2021high} provides a solution to proof the relations $\Cd\Td =\Cd$ and $\Td^{-1}\Bd = \Bd$.

In fact, the transformation algorithm presented in \cite[Section 4.1]{niederwieser2021high} can be drastically simplified since subsystem~(d) already exhibits a special structure.
The decomposition of the dynamic matrix in Step 1, equation (24) of the algorithm is straightforward as argued in the following.
The dynamic matrix of subsystem (d) obtained from any transformation to the SCB is given by
\begin{equation}
	\Adbar = \Ad^\star + \Bd \Fddbar + \Hddbar \Cd
\end{equation}
with $\Fddbar$ and $\Hddbar$ as matrices of appropriate dimensions. 
Using the notation of~\cite{niederwieser2021high}, this matrix is decomposed according to
\begin{equation}
\Adbar = \Adbarcheck - {\mt{\Pi}}\Cd,
\end{equation}
where $\Adbarcheck = \Ad^\star + \Bd \Fddbar$ and $\boldsymbol{\Pi} = - \Hddbar$.
Furthermore, an additional output transformation is not necessary, i.e., $\check \C = \Cd$ and $\mt{\Gamma} = \I_{\md}$.
The orders of the subsystems (denoted as $\mu_j$ in~\cite{niederwieser2021high}) are already given in sorted order by the lengths ${q_1 \geq q_2 \geq \cdots \geq q_{\md}}$ of the integrator chains. Then, the transformation algorithm yields the output matrix $\Cd =\diag{\C_{q_1},\C_{q_2},\ldots,\C_{q_{\md}}}$ for the transformed system which is ensured by \cite[Lemma 4.1.f)]{niederwieser2021high} and, thus, $\Cd \Td =\Cd$ holds.
In equation (B23) in the proof of \cite[Lemma 4.1.e]{niederwieser2021high}, it is shown that the input matrix of the transformed system is given by
\begin{equation}
\Td^{-1}\Bd = \boldsymbol{\mathcal{O}}_R \Bd.
\end{equation}
It can be easily shown that the reduced observability matrix
\begin{equation}
\boldsymbol{\mathcal{O}}_R = \I_{\nd}
\end{equation}
reduces to the identity matrix in this special case and, thus, $\Td^{-1}\Bd =\Bd$ is satisfied which completes the proof.

\section*{Author Biography}

\begin{biography}{\includegraphics[width=66pt,height=86pt]{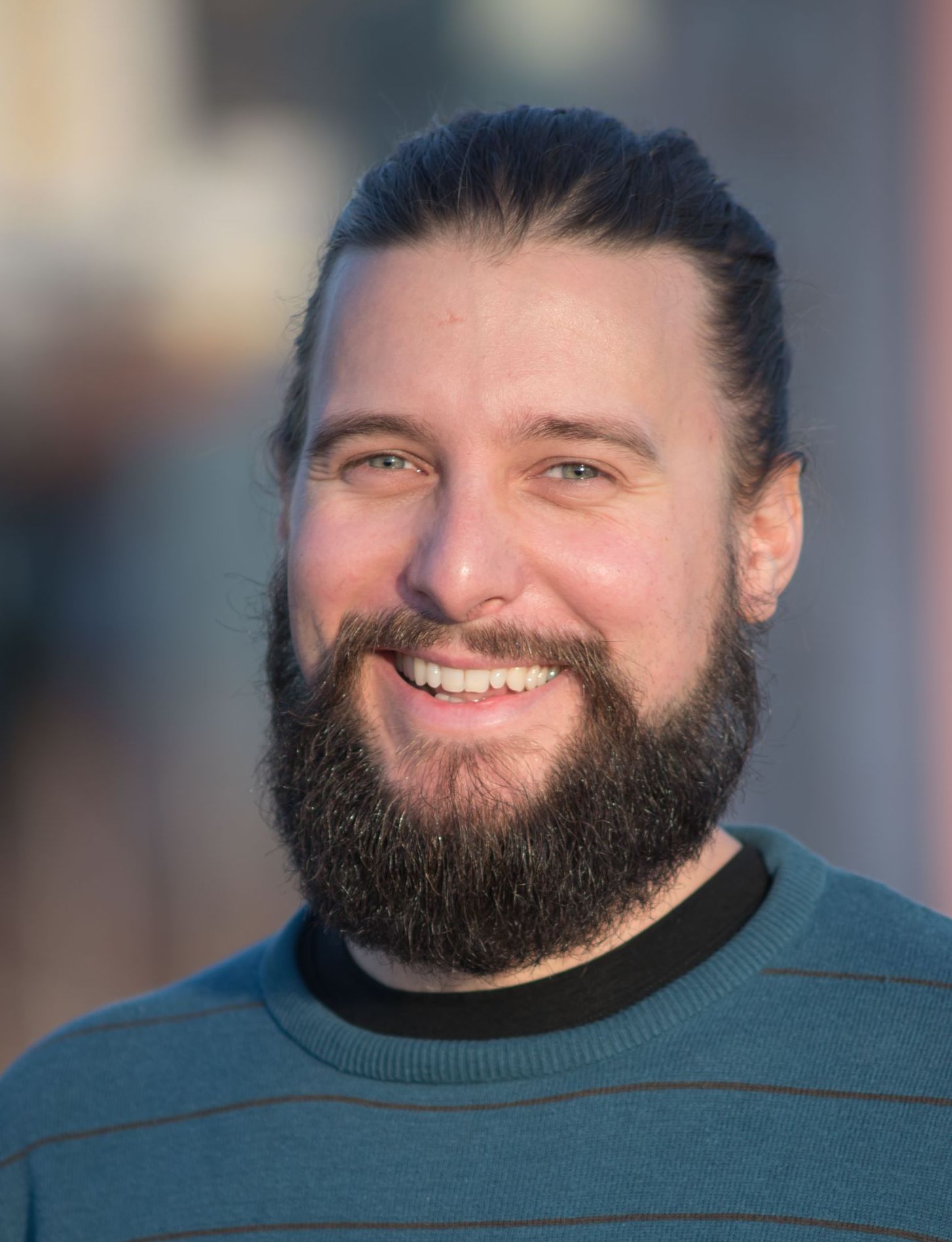}}{\textbf{Markus Tranninger}. Markus Tranninger received his M.Sc. degree in Electrical Engineering from Graz University of Technology in 2015, and he completed his Ph.D. at the Institute of Automation and Control, Graz University of Technology in 2020. He currently holds a Postdoc position at the Institute of Automation and Control at Graz University of Technology, Austria. He is part of the Graz University of Technology research center on Dependable Internet of Things. His research interests include state estimation and fault detection for complex dynamical systems.}
\end{biography}

\begin{biography}
	{\includegraphics[width=66pt,height=86pt]{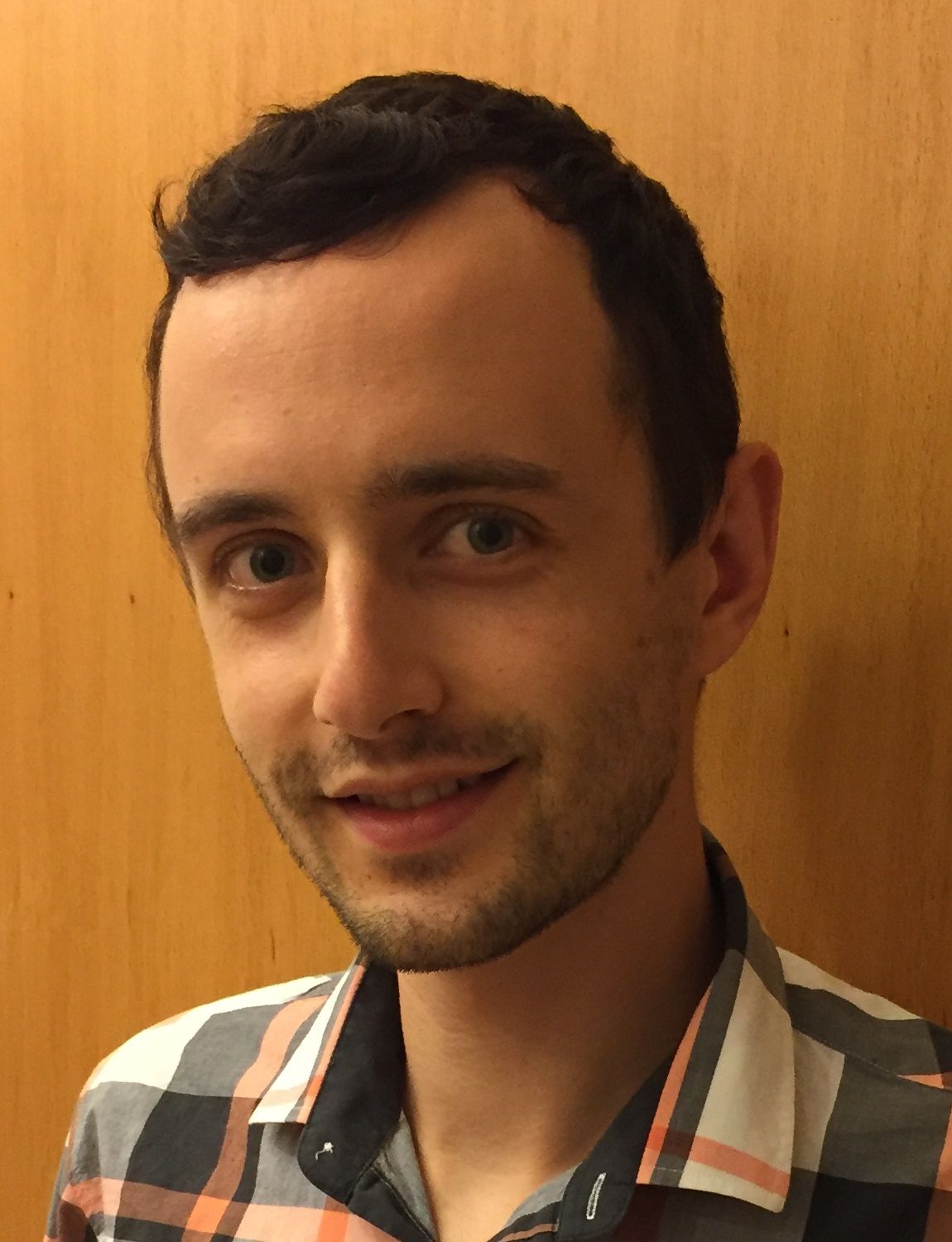}}{
		\textbf{Helmut Niederwieser} received his M.Sc. degree in Information and Computer Engineering from Graz University of Technology in 2019.
		He is a Ph.D. student at the Institute of Automation and Control, Graz University of Technology.
		He currently holds a Junior Researcher position at BEST – Bioenergy and Sustainable Technologies GmbH, Graz, Austria.
		His research interests include robust state and parameter estimation in thermochemical and thermotechnical processes.	
	}
\end{biography}

\begin{biography}{\includegraphics[width=66pt,height=86pt]{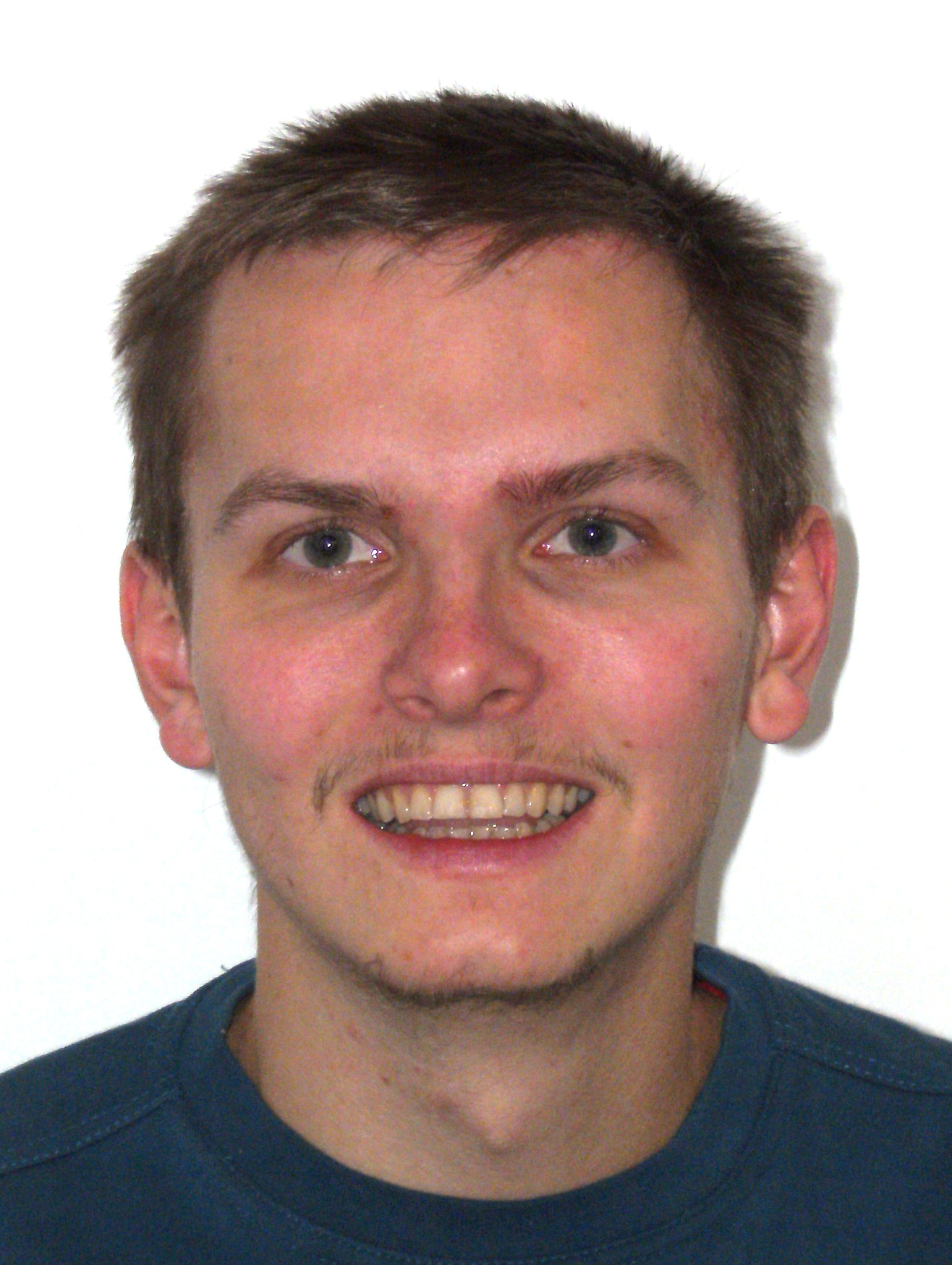}}{\textbf{Richard Seeber}. Richard Seeber received his M.Sc. degree in Electrical Engineering from Graz University of Technology in 2012, and he completed his Ph.D. at the Institute of Automation and Control, Graz University of Technology in 2017. He currently holds a Postdoc position at the Christian Doppler Laboratory for Model Based Control of Complex Test Bed Systems. His research interests include theory of sliding mode control systems, control of automotive test beds, and control of systems with actuator constraints.}
\end{biography}

\begin{biography}{\includegraphics[width=66pt,height=86pt]{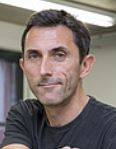}}{\textbf{Martin Horn} is professor and head of the Institute of Automation and Control at Graz University of Technology, Austria.  Until 2014 he was professor for control and measurement systems at Klagenfurt University, Austria. Since 2017  he is head of the Christian Doppler Laboratory for Model Based Control of Complex Textbed Systems. His research interests include robust and networked feedback systems with applications in automotive, semiconductor and pharmaceutical industry. }
\end{biography}

\end{document}